\documentclass[aps,showpacs,floatfix,twocolumn,byrevtex,superscriptaddress]{revtex4-1}

\usepackage{amsmath}
\usepackage{amssymb}
\usepackage{amstext}
\usepackage{amsopn}
\usepackage{amsfonts}
\usepackage{amsxtra}
\usepackage[multiple]{footmisc}
\usepackage{bookmark}
\usepackage[english]{babel}
\usepackage{amsmath}
\usepackage{graphicx}
\usepackage{float}
\usepackage{bm}
\usepackage{multirow}
\usepackage{dcolumn}
\usepackage[dvipsnames,usenames]{color}
\usepackage{hyperref}
\begin{document}

%%%%%%%%%%%%%%%%%%%%%%%%%%%%%%%%%%%%%%%%%%%%%%%%%%%%%%%%%%%%%%%%%%%%%
%%
%%  Title
%%

\title{Inherent circular dichroism of phonons in magnetic Weyl semimetal Co$_3$Sn$_2$S$_2$}

\author{R. Yang}
\thanks{These authors contributed equally to this work.}
\affiliation{Key Laboratory of Quantum Materials and Devices of Ministry of Education, School of Physics, Southeast University, Nanjing 211189, China}
\affiliation{1.~Physikalisches Institut Universit$\ddot{a}$t Stuttgart, 70569 Stuttgart, Germany}
\author{Y.-Y. Zhu}
\thanks{These authors contributed equally to this work.}
\affiliation{Key Laboratory of Quantum Materials and Devices of Ministry of Education, School of Physics, Southeast University, Nanjing 211189, China}
\author{M. Steigleder}
\affiliation{1.~Physikalisches Institut Universit$\ddot{a}$t Stuttgart, 70569 Stuttgart, Germany}
\author{X.-G. Qiu}
\affiliation{Institute of physics, Chinese academy of science, 100190 Beijing, China}
\author{T. -T. Zhang}
\email{ttzhang@itp.ac.cn}
\affiliation{Institute of theoretical physics, Chinese academy of science, 100190 Beijing, China}
\author{M. Dressel}
\email{dressel@pi1.physik.uni-stuttgart.de}
\affiliation{1.~Physikalisches Institut Universit$\ddot{a}$t Stuttgart, 70569 Stuttgart, Germany}
\date{\today}

%%%%%%%%%%%%%%%%%%%%%%%%%%%%%%%%%%%%%
%%
%% Abstract
%%
%

\begin{abstract}
We investigated the infrared-active phonons in ferromagnetic Weyl semimetal Co$_3$Sn$_2$S$_2$ using optical spectroscopy. Below the Curie temperature ($T_C\approx 175$~K), we observed asymmetric Fano lineshapes of phonons peaks in the optical conductivities, reflecting the presence of electron-phonon coupling (EPC). Additionally, the detected phonon signals by the polar Kerr rotation and the ellipticity spectroscopy indicate the circular dichroism (CD) of phonons. We attribute the CD of phonons to their distinct couplings with charge excitations on the tilted Weyl nodal rings in two circularly polarized channels. Our findings provide experimental evidence that, without external fields, phonons can also become circularly polarized by coupling with the electronic topology. Since the magnetic exchange splitting gradually shifts the topological bands in Co$_3$Sn$_2$S$_2$, the CD of phonons exhibits significant temperature dependence, hinting at a promising approach for manipulation.
\end{abstract}

%  72.15.-v  Electronic conduction in metals and alloys
%  74.70.-b  SC: Superconducting materials other than cuprates
%  78.20.-e  Optical properties of bulk materials and thin films
%  78.30.-j  Infrared and Raman spectra
%\pacs{72.15.-v, 74.70.-b, 78.30.-j}
\maketitle
%
%%%%%%%%%%%%%%%%%%%%%%%%%%%%%%%%%%%%%%%%%%%%%%%%%%%%%%%%%%%%%%%%%%%%%%%%%%%%%%%
% Introduction
%%%%%%%%%%%%%%%%%%%%%%%%%%%%%%%%%%%%%%%%%%%%%%%%%%%%%%%%%%%%%%%%%%%%%%%%%%%%%%%
%\section{Introduction}

\emph{Introduction.}---The chirality in bosonic system has garnered broad interest in condensed matter physics.
An intriguing example is the chiral phonon, which arises from the rotational movement of atoms perpendicular to their propagation direction~\cite{MARADUDIN1968, Streib2021,Ishito2023, Ueda2023,ZhangTT2023}.
Such circular motion imparts specific angular momentum and magnetic moment to the phonons~\cite{zhang2015chiral, Ren2021, Saparov2022}, giving rise to Haas effect, the phonon Hall effect, the anomalous thermal Hall effect, etc~\cite{Tauchert2022, Zhang2010, Kagan2008, Grissonnanche2019, Grissonnanche2020, Qin2012, Hu2021, Im2022, Kim2023, Juraschek2019}.
In these phenomena, circularly polarized phonons serve as carriers of energy and information~\cite{Yao2008, Tang2021, Chand2023, ZhangTT2018, Chen2019, ZhangTT2020, ZhangTT2022}.
Additionally, the circular motion of atoms in magnetic materials couples strongly with the electron spins, which can be used to manipulate the magnetism~\cite{Nova2017, Ren2024}.
Despite extensive theoretical investigations, circularly polarized phonons have only been observed in a limited range of materials~\cite{Zhu2018, Miao2018,Li2021,ZhangTT2023,Ishito2023,Ueda2023}.
Thus, realizing and manipulating circularly polarized phonons in more materials are crucial both for fundamental understanding and practical applications.

In the monolayer of the transition metal chalcogenides~\cite{Zhu2018}, circular modes were observed at $K$ and $K^{\prime}$ with zero propagating velocity; while in chiral crystals like $\alpha-$HgS~\cite{Ishito2023}, quartz~\cite{Ueda2023} and Te~\cite{ZhangTT2022, ZhangTT2023} chiral phonons with huge group velocity can be obtained in the vicinity of Brillouin zone center ($\Gamma$ point), due to the the non-centrosymmetric nature of those materials.
In Dirac semimetal Cd$_3$As$_2$~\cite{Cheng2020} and quantum magnet CoTiO$_3$~\cite{Lujan2024}, the coupling between phonons and charge excitations resulted in circularly polarized phonons with exceedingly large magnetic moments~\cite{Hernandez2023, Wu2023, Schaack1976, Luo2023, Baydin2022}.
However, external magnetic field is required to break the degeneracy of phonons with opposite polarizations.
Whether intrinsic circularly polarized phonons can exist in centrosymmetric systems with spontaneously $\mathcal{T}$-breaking remains an open question.
Very recently, theoretical studies predicted that, in systems with broken $\mathcal{T}$, the coupling between phonons and electronic topology could lift the degeneracy of phonons at the $\Gamma$ point, generating circularly polarized modes even without external magnetic field~\cite{Liu2017, Bistoni2021, Ren2021, Saparov2022}.
Nevertheless, experiments on such phenomenon in magnetic topological materials are still pending verification.
To reveal the circularly polarized phonons in magnetic topological materials, one needs to determine: (i) whether there exists considerable EPC, and (ii) whether the EPC can result in the CD of phonons.

%%%%%%%%%%%%%%%%%%%%%%%%%%%%%%%%%%%%%%%%%%%%%%%%%%%%%%%%%%%%%%%%%%%%%%%%%%%%%%%
% Figure 1 unpolarized optical conductivity
%%%%%%%%%%%%%%%%%%%%%%%%%%%%%%%%%%%%%%%%%%%%%%%%%%%%%%%%%%%%%%%%%%%%%%%%%%%%%%%
\begin{figure*}[tb]
    \centering
    \includegraphics [width=2\columnwidth]{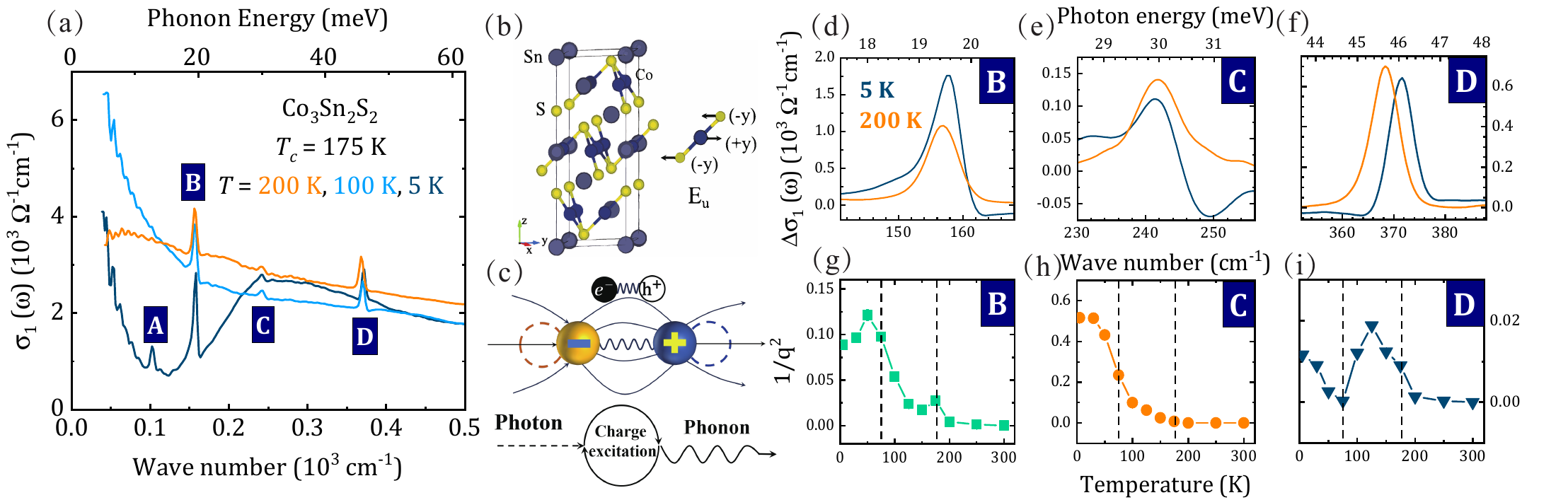}

\caption{Electron-phonon coupling below the Curie temperature. (a) The optical conductivity $\sigma_{1}(\omega)$ of Co$_3$Sn$_2$S$_2$ below 62~meV. The infrared-active phonons are marked by capital letters. (b)~Lattice structure (left) and the vibration of $E_u$ mode (right). (c) The cartoon (upper) and the diagram (lower)describing the EPC. (d-f)~The phonon lineshapes below and above the Curie temperature (175~K). The electronic background electronic background is subtracted
by polynomial fitting in narrow energy range. (g-i)~The Fano factors ($1/q^2$) acquired by fit the phonon lineshapes with Fano mode.}
\label{fig:optical spectroscopy}
\end{figure*}

Optical spectroscopy is a powerful tool for detecting the infrared-active phonons around the $\Gamma$ point.
The coupling between phonon and electronic systems can result in asymmetric Fano lineshape of the phonon absorption peaks in the optical conductivity, $\sigma_1(\omega)$~\cite{Yang2017, Xu2017, Cappelluti2012}.
On the other side, since circularly polarized phonons absorb right/left-circularly polarized light differently~\cite{Yao2008}, when the $\mathcal{T}$ of phonon field is broken, discernible CD can be reflected by finite polar Kerr rotation angle and ellipticity~\cite{Okamura2020, Hernandez2023, Levallois2015}.

In this work, with the reflectivity and polar Kerr rotation spectroscopy ($\theta_k(\omega)$) measurements, we studied the dynamic behaviors of the infrared-active phonons in magnetic Weyl semimetal Co$_3$Sn$_2$S$_2$, which has the centrosymmetric lattice structure (Fig.~\ref{fig:optical spectroscopy}b).
Below the $T_C$, Co$_3$Sn$_2$S$_2$ has six symmetry-related Weyl nodal rings without considering the spin-orbit coupling (SOC) and becomes a Weyl semimetal after considering the SOC.
In the far-infrared we observed asymmetric absorption peaks of the $E_u$ mode phonons, which has been regarded as the signature of the EPC.
The asymmetry of the phonon lineshapes become more prominent upon cooling, which suggests an increase in the coupling strength between phonons and charge excitations on the tilted Weyl nodal rings.
Meanwhile, the polar Kerr rotation and ellipticity spectroscopy detected significant CD of these phonons.
By analyzing the asymmetry of the phonon lineshapes in circularly polarized optical conductivities [$\sigma^{\pm}_{1}(\omega)$, $+/-$ represent right/left hand channel], we found that the phonons couple differently with the electronic topology for two circular polarizations.
Such unequal EPC in two polarizations breaks the $\mathcal{T}$ of phonon field, giving rise to the splitting of $E_u$ mode.
Moreover, since the magnetism gradually modulates the topological states, the EPC as well as the CD of the phonons also exhibits noticeable temperature ($T$) dependence.

%\section{Results}

%%%%%%%%%%%%%%%%%%%%%%%%%%%%%%%%%%%%%%%%%%%%%%%%%%%%%%%%%%%%%%%%%%%%%
% A . EPC in unpolarized optical conductivity

\emph{Electron-phonon coupling below the $T_C$.}---Fig.~\ref{fig:optical spectroscopy}(a) displays $\sigma_1(\omega)$ above and below the $T_{C}$ (see Sec.~I of the Supplemental Material (SM)~\footnote{See Supplemental Material at http://link.aps.org/supplemental for the sample characterizations and experimental techniques as well as for complementary data and analysis, which include Refs.~\onlinecite{Yang2020, Okamura2020, Levallois2015, Soh2022, Li2012, Cappelluti2012, Ren2021, Saparov2022}} and Ref.~\onlinecite{Yang2020} for details of the measurements).
The optical conductivity is dominated by the smooth and continuous responses from the absorptions for charge excitations~\cite{Yang2020, XuYS2020}, while several sharp peaks (marked by A, B, C and D) stem from the absorptions of infrared-active phonons.
Based on the first-principles calculations, these sharp peaks can be attributed to the $E_u$ mode vibrations (Fig.~S5 of the SM~\footnotemark[1]).
In this work, we will focus on the behavior of these phonons.
Upon cooling, along with the dramatic change of charge excitations, the phonon peaks exhibit a remarkable $T$ dependence.
In addition to the usual sharpening, their lineshapes become asymmetric as $T$ is reduced.
In Figs.~\ref{fig:optical spectroscopy}(d-f), after subtracting the electronic background using quadratic fit in short energy ranges, it becomes evident that the phonon lineshapes exhibit a remarkable asymmetry at low $T$.
Such asymmetric lineshape has been observed in graphene~\cite{Li2012}, superconductors~\cite{Yang2017, Xu2019}, and Weyl semimetals~\cite{Xu2017}, and attributed to the EPC, during which the discrete phonon excitations have comparable energies with the continuous charge excitations and interfere with each other~\cite{Cappelluti2012}.
The stronger EPC results in more asymmetric lineshape.
Since phonon A is screened by itinerant carriers above 75~K and exhibits a symmetric lineshape (Fig.~\ref{fig:optical spectroscopy}a), we will concentrate on the properties of phonons B, C, and D.
%%%%%%%%%%%%%%%%%%%%%%%%%%%%%%%%%%%%%%%%%%%%%%%%%%%%%%%%%%%%%%%%%%%%%%%%%%%%%%%%%%%%%%%%%%%%%%
%Figure 2:Circular dichroism of the phonons
%%%%%%%%%%%%%%%%%%%%%%%%%%%%%%%%%%%%%%%%%%%%%%%%%%%%%%%%%%%%%%%%%%%%%%%%%%%%%%%%%%%%%%%%%%%%%
\begin{figure*}[tb]
\centering
\includegraphics[width=1.5\columnwidth]{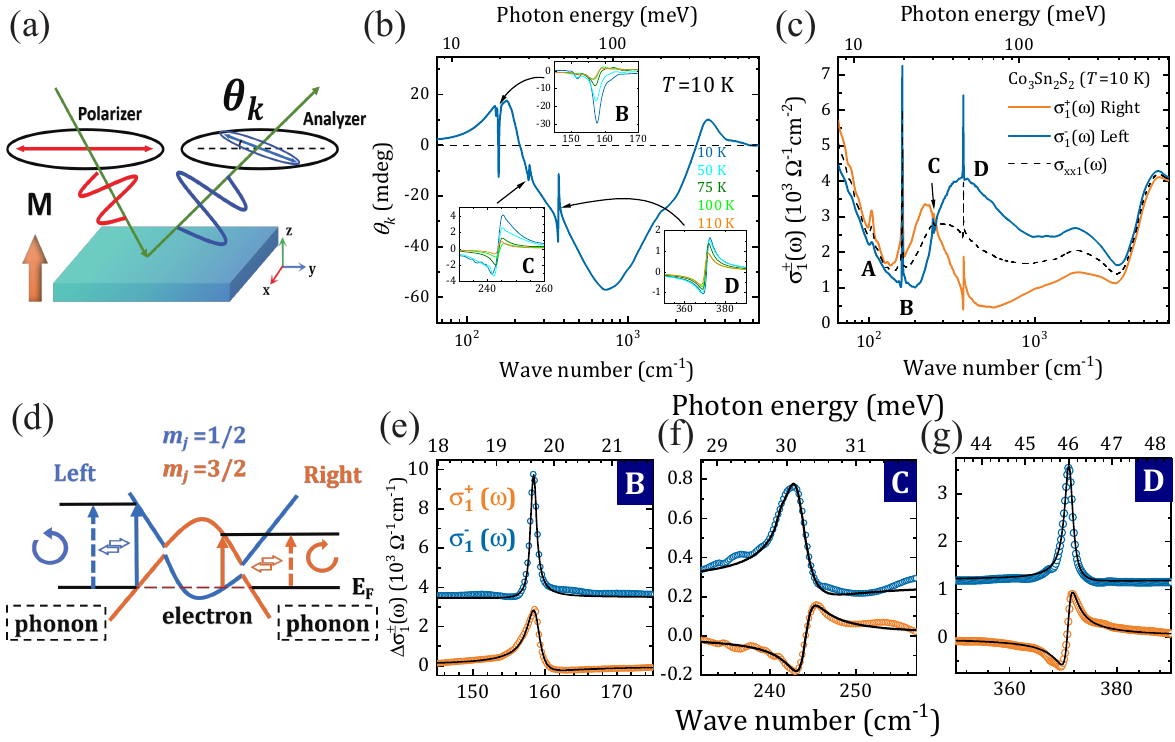}
\caption{Circular dichroism of the phonons in Co$_3$Sn$_2$S$_2$. (a) Sketch of the polar Kerr rotation measurement, during which the light is parallel to the magnetic moment but perpendicular to the sample surface. (b) The Kerr rotation angle spectroscopy below 0.8~eV measured at $T=10$~K. The insets depict the Kerr rotation spectroscopy of phonons B, C, and D measured at various $T$, in which the electronic background is subtracted. (c)~The circularly polarized optical conductivity $\sigma_{1}^{\pm}(\omega)$ of Co$_3$Sn$_2$S$_2$ at 10~K. The four infrared-active phonons are marked by the capital letters. The dashed line represents the unpolarized optical conductivity $\sigma_{xx1}(\omega)=1/2(\sigma_{1}^{+}(\omega)+\sigma_{1}^{-}(\omega))$. (d)~Sketch of EPC in each circular polarization, in which the bands constituting the Weyl nodal rings harbor different angular momenta of +3/2 (orange) and +1/2 (blue), respectively. The solid and dashed lines represent the direct interband transitions for phonons and electrons. (e-g) Lineshapes of phonons B, C, and D, in each circular polarization at 10~K. The black solid lines are fits by Fano model. }
\label{fig:circular dichroism}
\end{figure*}

The asymmetry of the phonon peaks can be quantitatively described by the Fano model,
\begin{equation}\label{Fano model}
  \sigma_1(\omega)=\frac{2\pi}{Z_0}\frac{\Omega^2}{\gamma}\frac{q^{2}+\frac{4q(\omega-\omega_0)}{\gamma}-1}{q^{2}(1+\frac{4(\omega-\omega_0)^{2}}{\gamma^{2}})},
\end{equation}
in which $Z_0$ is the vacuum impedance; $\omega_0$, $\gamma$, and $\Omega$ correspond to the resonance frequency, linewidth, and strength of the phonon, respectively; the Fano factor $1/q^2$ or $1/q$ quantifies the asymmetry, with a higher value indicating a more pronounced asymmetric lineshape (see Sec.~II of the SM~\footnotemark[1] for detail).
The $T$ dependence of $1/q^2$ displayed in Figs.~\ref{fig:optical spectroscopy}(g-i) reveals that the asymmetry of the phonons B, C, and D emerges right below $T_C\approx 175$~K and become more apparent at lower $T$, suggesting a crucial role played by magnetism in driving the EPC in Co$_3$Sn$_2$S$_2$.
While the asymmetry of the phonons B and C increases continuously upon cooling, phonon D asymmetry shows a non-monotonic $T$ dependence and reaches its minimum at 75~K, indicating two distinct EPC process, as discussed later.

At low $T$, the electronic background in $\sigma_1(\omega)$ below 0.1~eV was attributed to the charge excitations on tilted Weyl nodal rings~\cite{Okamura2020, Yang2020}.
Thus, for comparable energies, phonons B, C, and D primarily interfere with the charge excitations within the topological states~\footnote{Since these phonons are infrared active at the Brillouin zone center, they interfere with the direct interband transition as shown in Fig.~2d, which do not show momentum change.}.
SOC leads to distinct angular momenta of the bands constituting Weyl nodal rings (as shown in Fig.~\ref{fig:circular dichroism}d and Fig.~S7).
For the conservation of angular momentum, charge excitations on different sides of the Weyl nodal rings should absorb photons with different chiralities~\cite{Okamura2020} (Fig.~\ref{fig:circular dichroism}d).
However, the Weyl nodal rings in Co$_3$Sn$_2$S$_2$ are tilted, for the Pauli blockade effect~\cite{Ma2017}, charge excitations on the Weyl nodal rings absorb the photons with different energies in each circular polarization (sketched in Fig.~\ref{fig:circular dichroism}d).
Because phonons couple to charge excitations, it is intriguing whether the electronic system can transmit the CD to phonons through EPC in centrosymmetric Co$_3$Sn$_2$S$_2$.

%%%%%%%%%%%%%%%%%%%%%%%%%%%%%%%%%%%%%%%%%%%%%%%%%%%%%%%%%%%%%%%%%%%%%
% B . Circular dichroism of the phonons
%%%%%%%%%%%%%%%%%%%%%%%%%%%%%%%%%%%%%%%%%%%%%%%%%%%%%%%%%%%%%%%%%%%%%%

\emph{Circular dichroism of the phonons in Co$_3$Sn$_2$S$_2$.}---With the optical configuration in Fig.~\ref{fig:circular dichroism}(a), we measured the polar Kerr rotation and ellipticity of Co$_3$Sn$_2$S$_2$ (see Sec.~I of the SM~\footnotemark[1] and Ref.~\onlinecite{Levallois2015} for the detail of the measurements).
The polar Kerr effect arises from the difference in absorption of light with opposite circular polarizations~\cite{Okamura2020, Levallois2015}.
From the results shown in Fig.~\ref{fig:circular dichroism}b, one can see considerable Kerr rotation angle ($\theta_{k}$) in broad energy ranges at 10~K (the ellipticity ($\eta_k$) is shown in Fig. S1~\footnotemark[1]).
At 90~meV, $\theta_k$ is as large as 60~mrad, such strong Kerr rotation was also observed by Y. Okamura \emph{et al.} and attributed to the presence of tilted Weyl nodal rings near the Fermi level~\cite{Okamura2020}.
Below 0.1~eV, both $\theta_{k}$ and $\eta_k$ change signs, indicating that the charge excitations of different energies on the tilted Weyl nodal rings have distinct preferences for one of the polarizations~\cite{Sugano2000}.
In addition to the electronic responses, we also observed several sharp peaks in $\theta_k(\omega)$ and the $\eta_k(\omega)$ (Fig. S1~\footnotemark[1]), these sharp peaks occur at the same energies as the phonons B, C, and D, hinting at the CD of phonon absorptions~\cite{Hernandez2023, Baydin2022}.
With increasing $T$, the signals of these phonons gradually diminish (insets of Fig.~\ref{fig:circular dichroism}b).

In order to investigate the origin of CD in Co$_3$Sn$_2$S$_2$, we derived $\sigma^{\pm}_{1}(\omega)$ based on the reflectivity and the Kerr rotation spectra (Fig.~\ref{fig:circular dichroism}c, refer to Sec.~I of the SM~\footnotemark[1] and Ref.~\onlinecite{Okamura2020} for details).
As expected, at 10~K $\sigma^{\pm}_{1}(\omega)$ exhibit a significant disparity in the two circular polarizations.
The absorption peak at 30~meV in unpolarized optical conductivity (represented by the dashed curve in Fig.~\ref{fig:circular dichroism}c) was attributed to charge excitations on tilted Weyl nodal rings~\cite{Yang2020}.
As  mentioned above, the charge excitations on different sides of the tilted Weyl nodal rings absorb photons differently for the two circular polarizations (Fig.~\ref{fig:circular dichroism}d), thus, the 30~meV absorption peak splits with different energies for $\sigma^{+}_{1}(\omega)$ and $\sigma^{-}_{1}(\omega)$ (depicted by the blue and orange curves in Fig.~\ref{fig:circular dichroism}c)~\footnote{Since there is no anisotropy within the Co$_3$Sn$_2$S$_2$ \emph{ab}-plane, $\sigma_{xx1}(\omega)$ is equal to $\sigma_{1}(\omega)$.}.
%Such splitting further reflects the chirality of topological state.

\begin{figure}[ht]
\centering
\includegraphics[width=0.85\columnwidth]{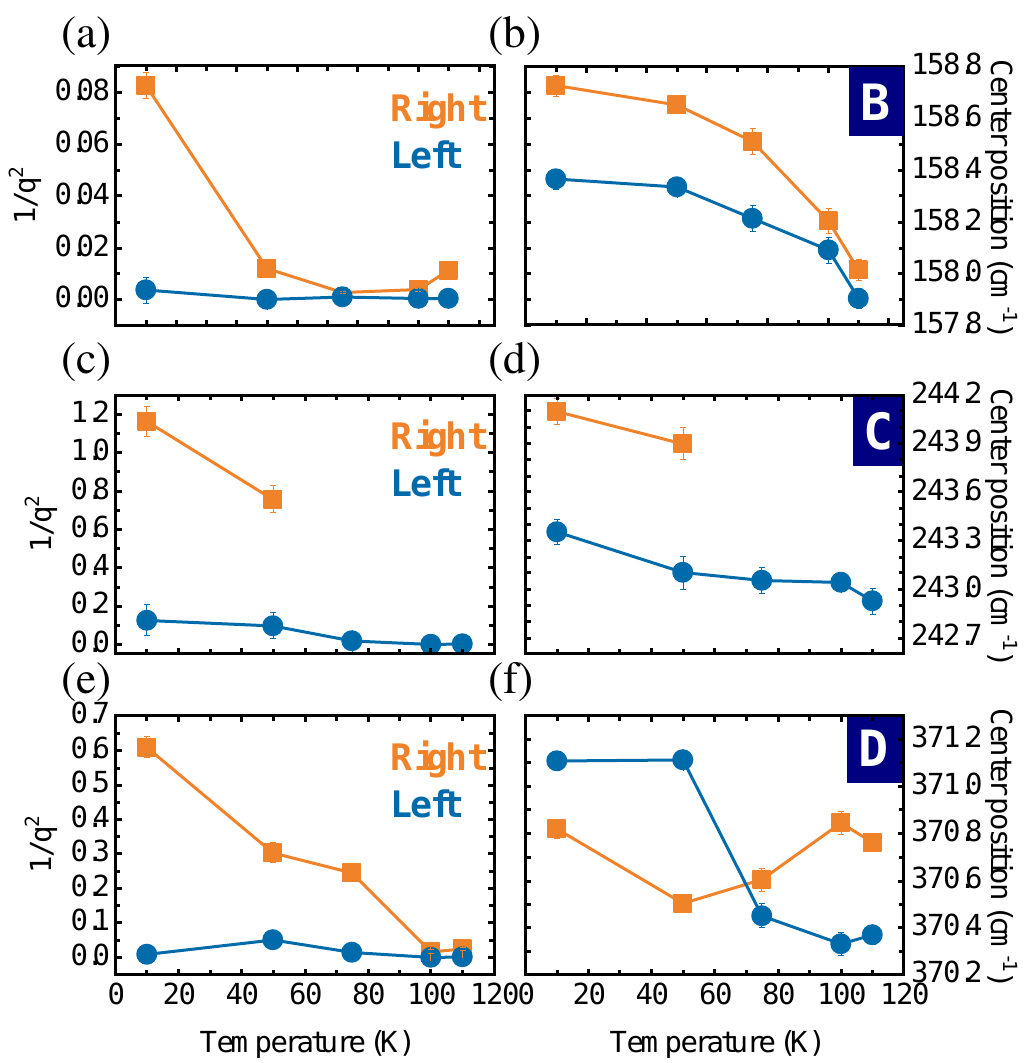}
\caption{Temperature dependence of the parameters used to fit the phonon modes B, C and D. (a,c,e)~Temperature dependence of the Fano factor $1/q^2$ in each circular polarization. (b,d,f) Temperature dependence of peak positions in each circular polarization.}
\label{fig:D-L fit}
\end{figure}

In $\sigma^{\pm}_{1}(\omega)$, after subtracting the electronic background, the phonon absorptions still show asymmetirc lineshapes.
However, the lineshapes for opposite polarizations are rather distinct for phonons B, C, and D, indicating different EPC (Figs.~\ref{fig:circular dichroism}e-g).
The phonon peaks are then fitted with the Fano model (black lines in Figs.~\ref{fig:circular dichroism}e-g and Fig.~S4).
Figures.~\ref{fig:D-L fit}a, c, and e clearly demonstrate that $ 1/q^2$ in the right-hand channel is significantly larger for phonons B, C, and D.
The distinct EPC also results in splitting of the phonon positions in two polarizations.
For phonons B and C, the center frequencies are higher in the right-hand channel, and the splitting between two circular polarizations increases as $T$ is lowered (Figs.~\ref{fig:D-L fit}b, d), while the splitting of phonon peak D exhibits a non-monotonic evolution that even reverses across 75~K (Fig.~\ref{fig:D-L fit}f).
Since $\sigma_1(\omega)$ is the average of the $\sigma^{\pm}_{1}(\omega)$, the reversal of phonon D splitting leads to the non-monotonic evolution of its asymmetry in Fig.~\ref{fig:optical spectroscopy}i.
The broken degeneracy of phonons imparts them with finite angular momentum~\cite{Saparov2022}.

%%%%%%%%%%%%%%%%%%%%%%%%%%%%%%%%%%%%%%%%%%%%%%%%%%%%%%%%%%%%%%%%%%%%%%%%%%%%%%%%%%%
%Figure 3 Fit parameters
%%%%%%%%%%%%%%%%%%%%%%%%%%%%%%%%%%%%%%%%%%%%%%%%%%%%%%%%%%%%%%%%%%%%%%%%%%%%%%%%%%%
%
%%%%%%%%%%%%%%%%%%%%%%%%%%%%%%%%%%%%%%%%%%%%%%%%%%%%%%%%%%%%%%%%%%%%%%%%%%%%%%%%%%%%%%%%%%%%%%%
%\section{Discussion}
%%%%%%%%%%%%%%%%%%%%%%%%%%%%%%%%%%%%%%%%%%%%%%%%%%%%%%%%%%%%%%%%%%%%%%%%%%%%%%%%%%%%%%%%%%%%%%%%
%
\emph{Discussion.}---So far, optical spectroscopy has revealed significant $T$-dependent EPC in magnetic Weyl semimetal Co$_3$Sn$_2$S$_2$ below $T_C$.
Since the energies of the infrared-active phonons overlap with those of charge excitations on the tilted Weyl nodal rings, the discrete phonons mainly interfere with the continuum of the topological electronic state.
Due to SOC and Pauli blockade effect, charge excitations on the tilted Weyl nodal rings exhibit remarkable CD.
A close inspection of $\sigma^{\pm}_{1}(\omega)$ reveals that the phonons B, C, and D couple differently with the charge excitations in two circular polarizations (Figs.~\ref{fig:circular dichroism}e-g).

The asymmetric phonon lineshape observed by optical spectroscopy is known as the charged-phonon effect: the phonon modes borrow dipole intensities from the electronic background (Fig.~\ref{fig:optical spectroscopy}c)~\cite{Cappelluti2012}.
In Co$_3$Sn$_2$S$_2$, the infrared-active phonons stem from linear dipole vibrations ($E_u$ mode), which can be regarded as the degeneracy of left- and right-handed circularly polarized modes.
As phonons exchange energies with the charge excitations distinctly in two circular polarizations, the degeneracy of phonons will be lifted.
Consequently, the linear vibrations transform into elliptical ones.

Moreover, the phonon-peak splitting in the two circular polarizations exhibit an interesting $T$ dependence.
For phonons B and C, the center frequency is always higher in the right-hand channel, with an expanding separation upon cooling (Figs.~\ref{fig:D-L fit}b and d).
The splitting of phonon D, however is rather different: when passing 75~K, it reverses the direction (Fig.~\ref{fig:D-L fit}f).
From theoretical considerations it was pointed out recently that in $\mathcal{T}$ breaking system, the molecular Berry curvature (MBC) induced by the coupling between phonons and electronic topology can break the degeneracy of the optical phonons at the $\Gamma$ point, resulting in phonon splitting into right- and left-hand polarized branches.
The direction and magnitude of the splitting are determined by the real part of MBC at the $\Gamma$ point~\cite{Saparov2022, Ren2021, Bistoni2021}.
In Ref.~\onlinecite{Saparov2022} EPC only happens in one circular polarization --~since the valleys are polarized,
leading to one splitting direction at the $\Gamma$ point.
However, the tilted Weyl nodal rings in Co$_3$Sn$_2$S$_2$ are formed by band crossing and therefore EPC occurs in both circular polarizations (Fig.~\ref{fig:circular dichroism}d); the resulting phonon splittings are opposite and compete with each other~\cite{Ren2021}.
Thus, the MBC should be considered in two circular polarizations.
According to its definition [Eq.~(17) of Ref.~\onlinecite{Saparov2022}], MBC at the $\Gamma$ point corresponds to the phonon-induced direct interband transitions \cite{Saparov2022}, which can be evaluated by the strength of EPC and the charge excitation probability.
The EPC can be quantized by the Fano factor $1/q$, while the probability of charge excitation is proportional to the electronic background in the optical conductivity $\sigma_{1}^{e\pm}(\omega_0)$.

In Fig.~S8 of the SM~\footnotemark[1], we plot the values of $|1/q\times\sigma_{1}^{e\pm}(\omega_0)|$, which corresponds to the MBC in opposite circular polarizations.
For phonons B and C, both the EPC and the charge excitations at the phonon frequencies $\omega_0$ are always stronger in right-hand polarization, resulting in larger MBC and higher phonon frequencies (Figs.~\ref{fig:D-L fit}b and d).
For phonon D, at $T<$ 75~K, even though the EPC is stronger in the right-hand polarization, the charge excitations are much stronger in the left-hand polarization (Fig.~\ref{fig:circular dichroism}c), giving rise to larger MBC and higher phonon frequencies for that polarization (Fig.~\ref{fig:D-L fit}f).
Since Co$_3$Sn$_2$S$_2$ is an itinerant ferromagnet, with increasing $T$ the magnetic exchange gradually shifts the Weyl nodal rings away from the Fermi level~\cite{Yang2020, Liu2021}.
With the suppressed charge excitations on the Weyl nodal rings and EPC in both circular polarizations, the splitting of the B and C phonons diminishes.
However, at the frequency of phonon D the gap in charge excitations between two circular polarizations narrows at high $T$ (Fig.~S9), then the larger EPC gives rise to larger MBC in right-hand polarization and reverses the splitting across 75~K.
The correspondence between $|1/q \times \sigma_{1}^{e\pm}(\omega_0)|$ and phonon splitting further reflects that the circular phonons stem from the couplings of phonons to the topological electronic states.

The reversal of phonon D splitting is reminiscent of the reversed phonon magnetic momentum or $g$ factor predicted~\cite{Ren2021} and observed in Pb$_{1-x}$Sn$_x$Te~\cite{Hernandez2023}; this phenomenon was attributed to the band inversion during the topological phase transitions.
Nevertheless, in these situations, EPC only occurs in one circular polarization and the external magnetic field is required to break the $\mathcal{T}$ of phonon field.
In Co$_3$Sn$_2$S$_2$, the EPC exists in both polarizations.
Due to the competition of two splitting effects, the Pauli blockade effect, and magnetism modification to the topological states, the properties of circular polarized phonons are especially rich here, even in the absence of external field.
In particular, the $T$-dependent phonon splitting in opposite circular polarizations provides an efficient way to manipulate the phonon polarization.
Since the magnetic exchange effect shifts the topological state~\cite{Yang2020, Liu2021}, manipulating the phonon polarization through doping~\cite{Neubauer2022} or gate voltage~\cite{XuQ2018} --~besides $T$~-- holds promise for future studies.

%%%%%%%%%%%%%%%%%%%%%%%%%%%%%%%%%%%%%%%%%%%%%%%%%%%%%%%%%%%%%%%%%%%%%%%%%%%%%%%%%%%%%%%%%%%%%%%%%%%%%%%%%%%%%%%%%%%%%%%%%%%%%%%
%\section{Conclusion}
%%%%%%%%%%%%%%%%%%%%%%%%%%%%%%%%%%%%%%%%%%%%%%%%%%%%%%%%%%%%%%%%%%%%%%%%%%%%%%%%%%%%%%%%%%%%%%%%%%%%%%%%%%%%%%%%%%%%%%%%%%%%%%%
\emph{Summary.}---Using the reflectivity and the polar Kerr rotation spectroscopy measurements, we firstly observed evidence of circularly polarized phonons in centrosymmetric magnetic Weyl semimetal Co$_3$Sn$_2$S$_2$ without external magnetic field.
By analyzing the asymmetric phonon lineshapes in the optical conductivity,
we attribute the CD of phonons to the unequal couplings between phonons and electronic topology in opposite circular polarizations.
As magnetism modifies the topological bands continuously, the CD of the phonons exhibit a pronounced $T$ dependence.
Thus, this proof-of-principle study establishes a new avenue for exploring circularly polarized phonons, and provides a more flexible approach to manipulate the circular polarization and even the chirality of phonons.

%%%%%%%%%%%%%%%%%%%%%%%%%%%%%%%%%%%%%%%%%%%%%%%%%%%%%%%%%%%%%%%%%%%%%%%%%%%%%%%
%
% Acknowledgment
%
\section{Acknowledgments}
We thank T. Zhang, A. V. Pronin, B. Xu, Y. -M Dai, and J. -H Zhou for fruitful discussions, and Gabriele Untereiner for the measurement support.
T. -T. Zhang acknowledges the support from the National Natural Science Foundation of China (Grant Nos. 12047503 and 12374165), and National Key R\&D Young Scientist Project 2023YFA1407400.
R.Yang acknowledges the support from the open research fund of Key Laboratory of Quantum Materials and Devices (Southeast University) Ministry of Education, the Alexander von Humboldt foundation, and the National Natural Science Foundation of China (Grant No. 12474152).
The project was funded by the Deutsche Forschungsgemeinschaft via DR228/51-3.
%
%
%
%
%%%%%%%%%%%%%%%%%%%%%%%%%%%%%%%%%%%%%%%%%%%%%%%%%%%%%%%%%%%%%%%%%%%%%%%%%%%%%%%
%
% Author contribution
%
\section{Author contribution}
R. Y. grew the single crystals and carried out the optical spectroscopy measurements. T. -T. Z. performed the first-principles calculations. R.Y. analyzed the data and prepared the manuscript with comments and suggestions from all authors. R. Y and M. D. supervised this project.
%
%\bibliography{CSS-CD}

\begin{thebibliography}{57}%
\makeatletter
\providecommand \@ifxundefined [1]{%
 \@ifx{#1\undefined}
}%
\providecommand \@ifnum [1]{%
 \ifnum #1\expandafter \@firstoftwo
 \else \expandafter \@secondoftwo
 \fi
}%
\providecommand \@ifx [1]{%
 \ifx #1\expandafter \@firstoftwo
 \else \expandafter \@secondoftwo
 \fi
}%
\providecommand \natexlab [1]{#1}%
\providecommand \enquote  [1]{``#1''}%
\providecommand \bibnamefont  [1]{#1}%
\providecommand \bibfnamefont [1]{#1}%
\providecommand \citenamefont [1]{#1}%
\providecommand \href@noop [0]{\@secondoftwo}%
\providecommand \href [0]{\begingroup \@sanitize@url \@href}%
\providecommand \@href[1]{\@@startlink{#1}\@@href}%
\providecommand \@@href[1]{\endgroup#1\@@endlink}%
\providecommand \@sanitize@url [0]{\catcode `\\12\catcode `\$12\catcode
  `\&12\catcode `\#12\catcode `\^12\catcode `\_12\catcode `\%12\relax}%
\providecommand \@@startlink[1]{}%
\providecommand \@@endlink[0]{}%
\providecommand \url  [0]{\begingroup\@sanitize@url \@url }%
\providecommand \@url [1]{\endgroup\@href {#1}{\urlprefix }}%
\providecommand \urlprefix  [0]{URL }%
\providecommand \Eprint [0]{\href }%
\providecommand \doibase [0]{http://dx.doi.org/}%
\providecommand \selectlanguage [0]{\@gobble}%
\providecommand \bibinfo  [0]{\@secondoftwo}%
\providecommand \bibfield  [0]{\@secondoftwo}%
\providecommand \translation [1]{[#1]}%
\providecommand \BibitemOpen [0]{}%
\providecommand \bibitemStop [0]{}%
\providecommand \bibitemNoStop [0]{.\EOS\space}%
\providecommand \EOS [0]{\spacefactor3000\relax}%
\providecommand \BibitemShut  [1]{\csname bibitem#1\endcsname}%
\let\auto@bib@innerbib\@empty
%</preamble>
\bibitem [{\citenamefont {MARADUDIN}\ and\ \citenamefont
  {VOSKO}(1968)}]{MARADUDIN1968}%
  \BibitemOpen
  \bibfield  {author} {\bibinfo {author} {\bibfnamefont {A.~A.}\ \bibnamefont
  {MARADUDIN}}\ and\ \bibinfo {author} {\bibfnamefont {S.~H.}\ \bibnamefont
  {VOSKO}},\ }\href {\doibase 10.1103/RevModPhys.40.1} {\bibfield  {journal}
  {\bibinfo  {journal} {Rev. Mod. Phys.}\ }\textbf {\bibinfo {volume} {40}},\
  \bibinfo {pages} {1} (\bibinfo {year} {1968})}\BibitemShut {NoStop}%
\bibitem [{\citenamefont {Streib}(2021)}]{Streib2021}%
  \BibitemOpen
  \bibfield  {author} {\bibinfo {author} {\bibfnamefont {S.}~\bibnamefont
  {Streib}},\ }\href {\doibase 10.1103/PhysRevB.103.L100409} {\bibfield
  {journal} {\bibinfo  {journal} {Phys. Rev. B}\ }\textbf {\bibinfo {volume}
  {103}},\ \bibinfo {pages} {L100409} (\bibinfo {year} {2021})}\BibitemShut
  {NoStop}%
\bibitem [{\citenamefont {Ishito}\ \emph {et~al.}(2023)\citenamefont {Ishito},
  \citenamefont {Mao}, \citenamefont {Kousaka}, \citenamefont {Togawa},
  \citenamefont {Iwasaki}, \citenamefont {Zhang}, \citenamefont {Murakami},
  \citenamefont {Kishine},\ and\ \citenamefont {Satoh}}]{Ishito2023}%
  \BibitemOpen
  \bibfield  {author} {\bibinfo {author} {\bibfnamefont {K.}~\bibnamefont
  {Ishito}}, \bibinfo {author} {\bibfnamefont {H.}~\bibnamefont {Mao}},
  \bibinfo {author} {\bibfnamefont {Y.}~\bibnamefont {Kousaka}}, \bibinfo
  {author} {\bibfnamefont {Y.}~\bibnamefont {Togawa}}, \bibinfo {author}
  {\bibfnamefont {S.}~\bibnamefont {Iwasaki}}, \bibinfo {author} {\bibfnamefont
  {T.}~\bibnamefont {Zhang}}, \bibinfo {author} {\bibfnamefont
  {S.}~\bibnamefont {Murakami}}, \bibinfo {author} {\bibfnamefont {J.-i.}\
  \bibnamefont {Kishine}}, \ and\ \bibinfo {author} {\bibfnamefont
  {T.}~\bibnamefont {Satoh}},\ }\href {\doibase 10.1038/s41567-022-01790-x}
  {\bibfield  {journal} {\bibinfo  {journal} {Nat. Phys.}\ }\textbf {\bibinfo
  {volume} {19}},\ \bibinfo {pages} {35} (\bibinfo {year} {2023})}\BibitemShut
  {NoStop}%
\bibitem [{\citenamefont {Ueda}\ \emph {et~al.}(2023)\citenamefont {Ueda},
  \citenamefont {Garcia-Fernandez}, \citenamefont {Agrestini}, \citenamefont
  {Romao}, \citenamefont {van~den Brink}, \citenamefont {Spaldin},
  \citenamefont {Zhou},\ and\ \citenamefont {Staub}}]{Ueda2023}%
  \BibitemOpen
  \bibfield  {author} {\bibinfo {author} {\bibfnamefont {H.}~\bibnamefont
  {Ueda}}, \bibinfo {author} {\bibfnamefont {M.}~\bibnamefont
  {Garcia-Fernandez}}, \bibinfo {author} {\bibfnamefont {S.}~\bibnamefont
  {Agrestini}}, \bibinfo {author} {\bibfnamefont {C.~P.}\ \bibnamefont
  {Romao}}, \bibinfo {author} {\bibfnamefont {J.}~\bibnamefont {van~den
  Brink}}, \bibinfo {author} {\bibfnamefont {N.~A.}\ \bibnamefont {Spaldin}},
  \bibinfo {author} {\bibfnamefont {K.~J.}\ \bibnamefont {Zhou}}, \ and\
  \bibinfo {author} {\bibfnamefont {U.}~\bibnamefont {Staub}},\ }\href
  {\doibase 10.1038/s41586-023-06016-5} {\bibfield  {journal} {\bibinfo
  {journal} {Nature}\ }\textbf {\bibinfo {volume} {618}},\ \bibinfo {pages}
  {946} (\bibinfo {year} {2023})}\BibitemShut {NoStop}%
\bibitem [{\citenamefont {Zhang}\ \emph {et~al.}(2023)\citenamefont {Zhang},
  \citenamefont {Huang}, \citenamefont {Pan}, \citenamefont {Du}, \citenamefont
  {Zhang},\ and\ \citenamefont {Murakami}}]{ZhangTT2023}%
  \BibitemOpen
  \bibfield  {author} {\bibinfo {author} {\bibfnamefont {T.}~\bibnamefont
  {Zhang}}, \bibinfo {author} {\bibfnamefont {Z.}~\bibnamefont {Huang}},
  \bibinfo {author} {\bibfnamefont {Z.}~\bibnamefont {Pan}}, \bibinfo {author}
  {\bibfnamefont {L.}~\bibnamefont {Du}}, \bibinfo {author} {\bibfnamefont
  {G.}~\bibnamefont {Zhang}}, \ and\ \bibinfo {author} {\bibfnamefont
  {S.}~\bibnamefont {Murakami}},\ }\href {\doibase
  10.1021/acs.nanolett.3c02132} {\bibfield  {journal} {\bibinfo  {journal}
  {Nano Lett.}\ }\textbf {\bibinfo {volume} {23}},\ \bibinfo {pages} {7561}
  (\bibinfo {year} {2023})}\BibitemShut {NoStop}%
\bibitem [{\citenamefont {Zhang}\ and\ \citenamefont
  {Niu}(2015)}]{zhang2015chiral}%
  \BibitemOpen
  \bibfield  {author} {\bibinfo {author} {\bibfnamefont {L.}~\bibnamefont
  {Zhang}}\ and\ \bibinfo {author} {\bibfnamefont {Q.}~\bibnamefont {Niu}},\
  }\href@noop {} {\bibfield  {journal} {\bibinfo  {journal} {Physical review
  letters}\ }\textbf {\bibinfo {volume} {115}},\ \bibinfo {pages} {115502}
  (\bibinfo {year} {2015})}\BibitemShut {NoStop}%
\bibitem [{\citenamefont {Ren}\ \emph {et~al.}(2021)\citenamefont {Ren},
  \citenamefont {Xiao}, \citenamefont {Saparov},\ and\ \citenamefont
  {Niu}}]{Ren2021}%
  \BibitemOpen
  \bibfield  {author} {\bibinfo {author} {\bibfnamefont {Y.}~\bibnamefont
  {Ren}}, \bibinfo {author} {\bibfnamefont {C.}~\bibnamefont {Xiao}}, \bibinfo
  {author} {\bibfnamefont {D.}~\bibnamefont {Saparov}}, \ and\ \bibinfo
  {author} {\bibfnamefont {Q.}~\bibnamefont {Niu}},\ }\href {\doibase
  10.1103/PhysRevLett.127.186403} {\bibfield  {journal} {\bibinfo  {journal}
  {Phys. Rev. Lett.}\ }\textbf {\bibinfo {volume} {127}},\ \bibinfo {pages}
  {186403} (\bibinfo {year} {2021})}\BibitemShut {NoStop}%
\bibitem [{\citenamefont {Saparov}\ \emph {et~al.}(2022)\citenamefont
  {Saparov}, \citenamefont {Xiong}, \citenamefont {Ren},\ and\ \citenamefont
  {Niu}}]{Saparov2022}%
  \BibitemOpen
  \bibfield  {author} {\bibinfo {author} {\bibfnamefont {D.}~\bibnamefont
  {Saparov}}, \bibinfo {author} {\bibfnamefont {B.}~\bibnamefont {Xiong}},
  \bibinfo {author} {\bibfnamefont {Y.}~\bibnamefont {Ren}}, \ and\ \bibinfo
  {author} {\bibfnamefont {Q.}~\bibnamefont {Niu}},\ }\href {\doibase
  10.1103/PhysRevB.105.064303} {\bibfield  {journal} {\bibinfo  {journal}
  {Phys. Rev. B}\ }\textbf {\bibinfo {volume} {105}},\ \bibinfo {pages}
  {064303} (\bibinfo {year} {2022})}\BibitemShut {NoStop}%
\bibitem [{\citenamefont {Tauchert}\ \emph {et~al.}(2022)\citenamefont
  {Tauchert}, \citenamefont {Volkov}, \citenamefont {Ehberger}, \citenamefont
  {Kazenwadel}, \citenamefont {Evers}, \citenamefont {Lange}, \citenamefont
  {Donges}, \citenamefont {Book}, \citenamefont {Kreuzpaintner}, \citenamefont
  {Nowak},\ and\ \citenamefont {Baum}}]{Tauchert2022}%
  \BibitemOpen
  \bibfield  {author} {\bibinfo {author} {\bibfnamefont {S.~R.}\ \bibnamefont
  {Tauchert}}, \bibinfo {author} {\bibfnamefont {M.}~\bibnamefont {Volkov}},
  \bibinfo {author} {\bibfnamefont {D.}~\bibnamefont {Ehberger}}, \bibinfo
  {author} {\bibfnamefont {D.}~\bibnamefont {Kazenwadel}}, \bibinfo {author}
  {\bibfnamefont {M.}~\bibnamefont {Evers}}, \bibinfo {author} {\bibfnamefont
  {H.}~\bibnamefont {Lange}}, \bibinfo {author} {\bibfnamefont
  {A.}~\bibnamefont {Donges}}, \bibinfo {author} {\bibfnamefont
  {A.}~\bibnamefont {Book}}, \bibinfo {author} {\bibfnamefont {W.}~\bibnamefont
  {Kreuzpaintner}}, \bibinfo {author} {\bibfnamefont {U.}~\bibnamefont
  {Nowak}}, \ and\ \bibinfo {author} {\bibfnamefont {P.}~\bibnamefont {Baum}},\
  }\href {\doibase 10.1038/s41586-021-04306-4} {\bibfield  {journal} {\bibinfo
  {journal} {Nature}\ }\textbf {\bibinfo {volume} {602}},\ \bibinfo {pages}
  {73} (\bibinfo {year} {2022})}\BibitemShut {NoStop}%
\bibitem [{\citenamefont {Zhang}\ \emph {et~al.}(2010)\citenamefont {Zhang},
  \citenamefont {Ren}, \citenamefont {Wang},\ and\ \citenamefont
  {Li}}]{Zhang2010}%
  \BibitemOpen
  \bibfield  {author} {\bibinfo {author} {\bibfnamefont {L.}~\bibnamefont
  {Zhang}}, \bibinfo {author} {\bibfnamefont {J.}~\bibnamefont {Ren}}, \bibinfo
  {author} {\bibfnamefont {J.-S.}\ \bibnamefont {Wang}}, \ and\ \bibinfo
  {author} {\bibfnamefont {B.}~\bibnamefont {Li}},\ }\href {\doibase
  10.1103/PhysRevLett.105.225901} {\bibfield  {journal} {\bibinfo  {journal}
  {Phys. Rev. Lett.}\ }\textbf {\bibinfo {volume} {105}},\ \bibinfo {pages}
  {225901} (\bibinfo {year} {2010})}\BibitemShut {NoStop}%
\bibitem [{\citenamefont {Kagan}\ and\ \citenamefont
  {Maksimov}(2008)}]{Kagan2008}%
  \BibitemOpen
  \bibfield  {author} {\bibinfo {author} {\bibfnamefont {Y.}~\bibnamefont
  {Kagan}}\ and\ \bibinfo {author} {\bibfnamefont {L.~A.}\ \bibnamefont
  {Maksimov}},\ }\href {\doibase 10.1103/PhysRevLett.100.145902} {\bibfield
  {journal} {\bibinfo  {journal} {Phys. Rev. Lett.}\ }\textbf {\bibinfo
  {volume} {100}},\ \bibinfo {pages} {145902} (\bibinfo {year}
  {2008})}\BibitemShut {NoStop}%
\bibitem [{\citenamefont {Grissonnanche}\ \emph {et~al.}(2019)\citenamefont
  {Grissonnanche}, \citenamefont {Legros}, \citenamefont {Badoux},
  \citenamefont {Lefran{\c{c}}ois}, \citenamefont {Zatko}, \citenamefont
  {Lizaire}, \citenamefont {Lalibert{\'{e}}}, \citenamefont {Gourgout},
  \citenamefont {Zhou}, \citenamefont {Pyon}, \citenamefont {Takayama},
  \citenamefont {Takagi}, \citenamefont {Ono}, \citenamefont {Doiron-Leyraud},\
  and\ \citenamefont {Taillefer}}]{Grissonnanche2019}%
  \BibitemOpen
  \bibfield  {author} {\bibinfo {author} {\bibfnamefont {G.}~\bibnamefont
  {Grissonnanche}}, \bibinfo {author} {\bibfnamefont {A.}~\bibnamefont
  {Legros}}, \bibinfo {author} {\bibfnamefont {S.}~\bibnamefont {Badoux}},
  \bibinfo {author} {\bibfnamefont {E.}~\bibnamefont {Lefran{\c{c}}ois}},
  \bibinfo {author} {\bibfnamefont {V.}~\bibnamefont {Zatko}}, \bibinfo
  {author} {\bibfnamefont {M.}~\bibnamefont {Lizaire}}, \bibinfo {author}
  {\bibfnamefont {F.}~\bibnamefont {Lalibert{\'{e}}}}, \bibinfo {author}
  {\bibfnamefont {A.}~\bibnamefont {Gourgout}}, \bibinfo {author}
  {\bibfnamefont {J.-S.}\ \bibnamefont {Zhou}}, \bibinfo {author}
  {\bibfnamefont {S.}~\bibnamefont {Pyon}}, \bibinfo {author} {\bibfnamefont
  {T.}~\bibnamefont {Takayama}}, \bibinfo {author} {\bibfnamefont
  {H.}~\bibnamefont {Takagi}}, \bibinfo {author} {\bibfnamefont
  {S.}~\bibnamefont {Ono}}, \bibinfo {author} {\bibfnamefont {N.}~\bibnamefont
  {Doiron-Leyraud}}, \ and\ \bibinfo {author} {\bibfnamefont {L.}~\bibnamefont
  {Taillefer}},\ }\href {\doibase 10.1038/s41586-019-1375-0} {\bibfield
  {journal} {\bibinfo  {journal} {Nature}\ }\textbf {\bibinfo {volume} {571}},\
  \bibinfo {pages} {376} (\bibinfo {year} {2019})}\BibitemShut {NoStop}%
\bibitem [{\citenamefont {Grissonnanche}\ \emph {et~al.}(2020)\citenamefont
  {Grissonnanche}, \citenamefont {Th{\'{e}}riault}, \citenamefont {Gourgout},
  \citenamefont {Boulanger}, \citenamefont {Lefran{\c{c}}ois}, \citenamefont
  {Ataei}, \citenamefont {Lalibert{\'{e}}}, \citenamefont {Dion}, \citenamefont
  {Zhou}, \citenamefont {Pyon}, \citenamefont {Takayama}, \citenamefont
  {Takagi}, \citenamefont {Doiron-Leyraud},\ and\ \citenamefont
  {Taillefer}}]{Grissonnanche2020}%
  \BibitemOpen
  \bibfield  {author} {\bibinfo {author} {\bibfnamefont {G.}~\bibnamefont
  {Grissonnanche}}, \bibinfo {author} {\bibfnamefont {S.}~\bibnamefont
  {Th{\'{e}}riault}}, \bibinfo {author} {\bibfnamefont {A.}~\bibnamefont
  {Gourgout}}, \bibinfo {author} {\bibfnamefont {M.-E.}\ \bibnamefont
  {Boulanger}}, \bibinfo {author} {\bibfnamefont {E.}~\bibnamefont
  {Lefran{\c{c}}ois}}, \bibinfo {author} {\bibfnamefont {A.}~\bibnamefont
  {Ataei}}, \bibinfo {author} {\bibfnamefont {F.}~\bibnamefont
  {Lalibert{\'{e}}}}, \bibinfo {author} {\bibfnamefont {M.}~\bibnamefont
  {Dion}}, \bibinfo {author} {\bibfnamefont {J.-S.}\ \bibnamefont {Zhou}},
  \bibinfo {author} {\bibfnamefont {S.}~\bibnamefont {Pyon}}, \bibinfo {author}
  {\bibfnamefont {T.}~\bibnamefont {Takayama}}, \bibinfo {author}
  {\bibfnamefont {H.}~\bibnamefont {Takagi}}, \bibinfo {author} {\bibfnamefont
  {N.}~\bibnamefont {Doiron-Leyraud}}, \ and\ \bibinfo {author} {\bibfnamefont
  {L.}~\bibnamefont {Taillefer}},\ }\href {\doibase 10.1038/s41567-020-0965-y}
  {\bibfield  {journal} {\bibinfo  {journal} {Nat. Phys.}\ }\textbf {\bibinfo
  {volume} {16}},\ \bibinfo {pages} {1108} (\bibinfo {year}
  {2020})}\BibitemShut {NoStop}%
\bibitem [{\citenamefont {Qin}\ \emph {et~al.}(2012)\citenamefont {Qin},
  \citenamefont {Zhou},\ and\ \citenamefont {Shi}}]{Qin2012}%
  \BibitemOpen
  \bibfield  {author} {\bibinfo {author} {\bibfnamefont {T.}~\bibnamefont
  {Qin}}, \bibinfo {author} {\bibfnamefont {J.}~\bibnamefont {Zhou}}, \ and\
  \bibinfo {author} {\bibfnamefont {J.}~\bibnamefont {Shi}},\ }\href {\doibase
  10.1103/PhysRevB.86.104305} {\bibfield  {journal} {\bibinfo  {journal} {Phys.
  Rev. B}\ }\textbf {\bibinfo {volume} {86}},\ \bibinfo {pages} {104305}
  (\bibinfo {year} {2012})}\BibitemShut {NoStop}%
\bibitem [{\citenamefont {Hu}\ \emph {et~al.}(2021)\citenamefont {Hu},
  \citenamefont {Yu}, \citenamefont {Garate},\ and\ \citenamefont
  {Liu}}]{Hu2021}%
  \BibitemOpen
  \bibfield  {author} {\bibinfo {author} {\bibfnamefont {L.-H.}\ \bibnamefont
  {Hu}}, \bibinfo {author} {\bibfnamefont {J.}~\bibnamefont {Yu}}, \bibinfo
  {author} {\bibfnamefont {I.}~\bibnamefont {Garate}}, \ and\ \bibinfo {author}
  {\bibfnamefont {C.-X.}\ \bibnamefont {Liu}},\ }\href {\doibase
  10.1103/PhysRevLett.127.125901} {\bibfield  {journal} {\bibinfo  {journal}
  {Phys. Rev. Lett.}\ }\textbf {\bibinfo {volume} {127}},\ \bibinfo {pages}
  {125901} (\bibinfo {year} {2021})}\BibitemShut {NoStop}%
\bibitem [{\citenamefont {Im}\ \emph {et~al.}(2022)\citenamefont {Im},
  \citenamefont {Kim},\ and\ \citenamefont {Jin}}]{Im2022}%
  \BibitemOpen
  \bibfield  {author} {\bibinfo {author} {\bibfnamefont {J.}~\bibnamefont
  {Im}}, \bibinfo {author} {\bibfnamefont {C.~H.}\ \bibnamefont {Kim}}, \ and\
  \bibinfo {author} {\bibfnamefont {H.}~\bibnamefont {Jin}},\ }\href {\doibase
  10.1021/acs.nanolett.2c03095} {\bibfield  {journal} {\bibinfo  {journal}
  {Nano Lett.}\ }\textbf {\bibinfo {volume} {22}},\ \bibinfo {pages} {8281}
  (\bibinfo {year} {2022})}\BibitemShut {NoStop}%
\bibitem [{\citenamefont {Kim}\ \emph {et~al.}(2023)\citenamefont {Kim},
  \citenamefont {Vetter}, \citenamefont {Yan}, \citenamefont {Yang},
  \citenamefont {Wang}, \citenamefont {Sun}, \citenamefont {Yang},
  \citenamefont {Comstock}, \citenamefont {Li}, \citenamefont {Zhou},
  \citenamefont {Zhang}, \citenamefont {You}, \citenamefont {Sun},\ and\
  \citenamefont {Liu}}]{Kim2023}%
  \BibitemOpen
  \bibfield  {author} {\bibinfo {author} {\bibfnamefont {K.}~\bibnamefont
  {Kim}}, \bibinfo {author} {\bibfnamefont {E.}~\bibnamefont {Vetter}},
  \bibinfo {author} {\bibfnamefont {L.}~\bibnamefont {Yan}}, \bibinfo {author}
  {\bibfnamefont {C.}~\bibnamefont {Yang}}, \bibinfo {author} {\bibfnamefont
  {Z.}~\bibnamefont {Wang}}, \bibinfo {author} {\bibfnamefont {R.}~\bibnamefont
  {Sun}}, \bibinfo {author} {\bibfnamefont {Y.}~\bibnamefont {Yang}}, \bibinfo
  {author} {\bibfnamefont {A.~H.}\ \bibnamefont {Comstock}}, \bibinfo {author}
  {\bibfnamefont {X.}~\bibnamefont {Li}}, \bibinfo {author} {\bibfnamefont
  {J.}~\bibnamefont {Zhou}}, \bibinfo {author} {\bibfnamefont {L.}~\bibnamefont
  {Zhang}}, \bibinfo {author} {\bibfnamefont {W.}~\bibnamefont {You}}, \bibinfo
  {author} {\bibfnamefont {D.}~\bibnamefont {Sun}}, \ and\ \bibinfo {author}
  {\bibfnamefont {J.}~\bibnamefont {Liu}},\ }\href {\doibase
  10.1038/s41563-023-01473-9} {\bibfield  {journal} {\bibinfo  {journal} {Nat.
  Mater.}\ }\textbf {\bibinfo {volume} {22}},\ \bibinfo {pages} {322} (\bibinfo
  {year} {2023})}\BibitemShut {NoStop}%
\bibitem [{\citenamefont {Juraschek}\ and\ \citenamefont
  {Spaldin}(2019)}]{Juraschek2019}%
  \BibitemOpen
  \bibfield  {author} {\bibinfo {author} {\bibfnamefont {D.~M.}\ \bibnamefont
  {Juraschek}}\ and\ \bibinfo {author} {\bibfnamefont {N.~A.}\ \bibnamefont
  {Spaldin}},\ }\href {\doibase 10.1103/PhysRevMaterials.3.064405} {\bibfield
  {journal} {\bibinfo  {journal} {Physical Review Materials}\ }\textbf
  {\bibinfo {volume} {3}},\ \bibinfo {pages} {064405} (\bibinfo {year}
  {2019})}\BibitemShut {NoStop}%
\bibitem [{\citenamefont {Yao}\ \emph {et~al.}(2008)\citenamefont {Yao},
  \citenamefont {Xiao},\ and\ \citenamefont {Niu}}]{Yao2008}%
  \BibitemOpen
  \bibfield  {author} {\bibinfo {author} {\bibfnamefont {W.}~\bibnamefont
  {Yao}}, \bibinfo {author} {\bibfnamefont {D.}~\bibnamefont {Xiao}}, \ and\
  \bibinfo {author} {\bibfnamefont {Q.}~\bibnamefont {Niu}},\ }\href {\doibase
  10.1103/PhysRevB.77.235406} {\bibfield  {journal} {\bibinfo  {journal} {Phys.
  Rev. B}\ }\textbf {\bibinfo {volume} {77}},\ \bibinfo {pages} {235406}
  (\bibinfo {year} {2008})}\BibitemShut {NoStop}%
\bibitem [{\citenamefont {Tang}\ and\ \citenamefont {Cao}(2021)}]{Tang2021}%
  \BibitemOpen
  \bibfield  {author} {\bibinfo {author} {\bibfnamefont {D.-S.}\ \bibnamefont
  {Tang}}\ and\ \bibinfo {author} {\bibfnamefont {B.-Y.}\ \bibnamefont {Cao}},\
  }\href {\doibase 10.1063/5.0043623} {\bibfield  {journal} {\bibinfo
  {journal} {Journal of Applied Physics}\ }\textbf {\bibinfo {volume} {129}}
  (\bibinfo {year} {2021}),\ 10.1063/5.0043623}\BibitemShut {NoStop}%
\bibitem [{\citenamefont {Chand}\ \emph {et~al.}(2023)\citenamefont {Chand},
  \citenamefont {Woods}, \citenamefont {Quan}, \citenamefont {Mejia},
  \citenamefont {Taniguchi}, \citenamefont {Watanabe}, \citenamefont
  {Al{\`{u}}},\ and\ \citenamefont {Grosso}}]{Chand2023}%
  \BibitemOpen
  \bibfield  {author} {\bibinfo {author} {\bibfnamefont {S.~B.}\ \bibnamefont
  {Chand}}, \bibinfo {author} {\bibfnamefont {J.~M.}\ \bibnamefont {Woods}},
  \bibinfo {author} {\bibfnamefont {J.}~\bibnamefont {Quan}}, \bibinfo {author}
  {\bibfnamefont {E.}~\bibnamefont {Mejia}}, \bibinfo {author} {\bibfnamefont
  {T.}~\bibnamefont {Taniguchi}}, \bibinfo {author} {\bibfnamefont
  {K.}~\bibnamefont {Watanabe}}, \bibinfo {author} {\bibfnamefont
  {A.}~\bibnamefont {Al{\`{u}}}}, \ and\ \bibinfo {author} {\bibfnamefont
  {G.}~\bibnamefont {Grosso}},\ }\href {\doibase 10.1038/s41467-023-39339-y}
  {\bibfield  {journal} {\bibinfo  {journal} {Nature Communications}\ }\textbf
  {\bibinfo {volume} {14}},\ \bibinfo {pages} {3712} (\bibinfo {year}
  {2023})}\BibitemShut {NoStop}%
\bibitem [{\citenamefont {Zhang}\ \emph {et~al.}(2018)\citenamefont {Zhang},
  \citenamefont {Song}, \citenamefont {Alexandradinata}, \citenamefont {Weng},
  \citenamefont {Fang}, \citenamefont {Lu},\ and\ \citenamefont
  {Fang}}]{ZhangTT2018}%
  \BibitemOpen
  \bibfield  {author} {\bibinfo {author} {\bibfnamefont {T.}~\bibnamefont
  {Zhang}}, \bibinfo {author} {\bibfnamefont {Z.}~\bibnamefont {Song}},
  \bibinfo {author} {\bibfnamefont {A.}~\bibnamefont {Alexandradinata}},
  \bibinfo {author} {\bibfnamefont {H.}~\bibnamefont {Weng}}, \bibinfo {author}
  {\bibfnamefont {C.}~\bibnamefont {Fang}}, \bibinfo {author} {\bibfnamefont
  {L.}~\bibnamefont {Lu}}, \ and\ \bibinfo {author} {\bibfnamefont
  {Z.}~\bibnamefont {Fang}},\ }\href {\doibase 10.1103/PhysRevLett.120.016401}
  {\bibfield  {journal} {\bibinfo  {journal} {Phys. Rev. Lett.}\ }\textbf
  {\bibinfo {volume} {120}},\ \bibinfo {pages} {016401} (\bibinfo {year}
  {2018})}\BibitemShut {NoStop}%
\bibitem [{\citenamefont {Chen}\ \emph {et~al.}(2019)\citenamefont {Chen},
  \citenamefont {Lu}, \citenamefont {Dubey}, \citenamefont {Yao}, \citenamefont
  {Liu}, \citenamefont {Wang}, \citenamefont {Xiong}, \citenamefont {Zhang},\
  and\ \citenamefont {Srivastava}}]{Chen2019}%
  \BibitemOpen
  \bibfield  {author} {\bibinfo {author} {\bibfnamefont {X.}~\bibnamefont
  {Chen}}, \bibinfo {author} {\bibfnamefont {X.}~\bibnamefont {Lu}}, \bibinfo
  {author} {\bibfnamefont {S.}~\bibnamefont {Dubey}}, \bibinfo {author}
  {\bibfnamefont {Q.}~\bibnamefont {Yao}}, \bibinfo {author} {\bibfnamefont
  {S.}~\bibnamefont {Liu}}, \bibinfo {author} {\bibfnamefont {X.}~\bibnamefont
  {Wang}}, \bibinfo {author} {\bibfnamefont {Q.}~\bibnamefont {Xiong}},
  \bibinfo {author} {\bibfnamefont {L.}~\bibnamefont {Zhang}}, \ and\ \bibinfo
  {author} {\bibfnamefont {A.}~\bibnamefont {Srivastava}},\ }\href {\doibase
  10.1038/s41567-018-0366-7} {\bibfield  {journal} {\bibinfo  {journal} {Nat.
  Phys.}\ }\textbf {\bibinfo {volume} {15}},\ \bibinfo {pages} {221} (\bibinfo
  {year} {2019})}\BibitemShut {NoStop}%
\bibitem [{\citenamefont {Zhang}\ \emph {et~al.}(2020)\citenamefont {Zhang},
  \citenamefont {Takahashi}, \citenamefont {Fang},\ and\ \citenamefont
  {Murakami}}]{ZhangTT2020}%
  \BibitemOpen
  \bibfield  {author} {\bibinfo {author} {\bibfnamefont {T.}~\bibnamefont
  {Zhang}}, \bibinfo {author} {\bibfnamefont {R.}~\bibnamefont {Takahashi}},
  \bibinfo {author} {\bibfnamefont {C.}~\bibnamefont {Fang}}, \ and\ \bibinfo
  {author} {\bibfnamefont {S.}~\bibnamefont {Murakami}},\ }\href {\doibase
  10.1103/PhysRevB.102.125148} {\bibfield  {journal} {\bibinfo  {journal}
  {Phys. Rev. B}\ }\textbf {\bibinfo {volume} {102}},\ \bibinfo {pages}
  {125148} (\bibinfo {year} {2020})}\BibitemShut {NoStop}%
\bibitem [{\citenamefont {Zhang}\ and\ \citenamefont
  {Murakami}(2022)}]{ZhangTT2022}%
  \BibitemOpen
  \bibfield  {author} {\bibinfo {author} {\bibfnamefont {T.}~\bibnamefont
  {Zhang}}\ and\ \bibinfo {author} {\bibfnamefont {S.}~\bibnamefont
  {Murakami}},\ }\href {\doibase 10.1103/PhysRevResearch.4.L012024} {\bibfield
  {journal} {\bibinfo  {journal} {Phys. Rev. Research}\ }\textbf {\bibinfo
  {volume} {4}} (\bibinfo {year} {2022}),\
  10.1103/PhysRevResearch.4.L012024}\BibitemShut {NoStop}%
\bibitem [{\citenamefont {Nova}\ \emph {et~al.}(2017)\citenamefont {Nova},
  \citenamefont {Cartella}, \citenamefont {Cantaluppi}, \citenamefont
  {F{\"{o}}rst}, \citenamefont {Bossini}, \citenamefont {Mikhaylovskiy},
  \citenamefont {Kimel}, \citenamefont {Merlin},\ and\ \citenamefont
  {Cavalleri}}]{Nova2017}%
  \BibitemOpen
  \bibfield  {author} {\bibinfo {author} {\bibfnamefont {T.~F.}\ \bibnamefont
  {Nova}}, \bibinfo {author} {\bibfnamefont {A.}~\bibnamefont {Cartella}},
  \bibinfo {author} {\bibfnamefont {A.}~\bibnamefont {Cantaluppi}}, \bibinfo
  {author} {\bibfnamefont {M.}~\bibnamefont {F{\"{o}}rst}}, \bibinfo {author}
  {\bibfnamefont {D.}~\bibnamefont {Bossini}}, \bibinfo {author} {\bibfnamefont
  {R.~V.}\ \bibnamefont {Mikhaylovskiy}}, \bibinfo {author} {\bibfnamefont
  {A.~V.}\ \bibnamefont {Kimel}}, \bibinfo {author} {\bibfnamefont
  {R.}~\bibnamefont {Merlin}}, \ and\ \bibinfo {author} {\bibfnamefont
  {A.}~\bibnamefont {Cavalleri}},\ }\href {\doibase 10.1038/nphys3925}
  {\bibfield  {journal} {\bibinfo  {journal} {Nat. Phys.}\ }\textbf {\bibinfo
  {volume} {13}},\ \bibinfo {pages} {132} (\bibinfo {year} {2017})}\BibitemShut
  {NoStop}%
\bibitem [{\citenamefont {Ren}\ \emph {et~al.}(2024)\citenamefont {Ren},
  \citenamefont {Rudner},\ and\ \citenamefont {Xiao}}]{Ren2024}%
  \BibitemOpen
  \bibfield  {author} {\bibinfo {author} {\bibfnamefont {Y.}~\bibnamefont
  {Ren}}, \bibinfo {author} {\bibfnamefont {M.}~\bibnamefont {Rudner}}, \ and\
  \bibinfo {author} {\bibfnamefont {D.}~\bibnamefont {Xiao}},\ }\href {\doibase
  10.1103/PhysRevLett.132.096702} {\bibfield  {journal} {\bibinfo  {journal}
  {Phys. Rev. Lett.}\ }\textbf {\bibinfo {volume} {132}},\ \bibinfo {pages}
  {096702} (\bibinfo {year} {2024})}\BibitemShut {NoStop}%
\bibitem [{\citenamefont {Zhu}\ \emph {et~al.}(2018)\citenamefont {Zhu},
  \citenamefont {Yi}, \citenamefont {Li}, \citenamefont {Xiao}, \citenamefont
  {Zhang}, \citenamefont {Yang}, \citenamefont {Kaindl}, \citenamefont {Li},
  \citenamefont {Wang},\ and\ \citenamefont {Zhang}}]{Zhu2018}%
  \BibitemOpen
  \bibfield  {author} {\bibinfo {author} {\bibfnamefont {H.}~\bibnamefont
  {Zhu}}, \bibinfo {author} {\bibfnamefont {J.}~\bibnamefont {Yi}}, \bibinfo
  {author} {\bibfnamefont {M.~Y.}\ \bibnamefont {Li}}, \bibinfo {author}
  {\bibfnamefont {J.}~\bibnamefont {Xiao}}, \bibinfo {author} {\bibfnamefont
  {L.}~\bibnamefont {Zhang}}, \bibinfo {author} {\bibfnamefont {C.~W.}\
  \bibnamefont {Yang}}, \bibinfo {author} {\bibfnamefont {R.~A.}\ \bibnamefont
  {Kaindl}}, \bibinfo {author} {\bibfnamefont {L.~J.}\ \bibnamefont {Li}},
  \bibinfo {author} {\bibfnamefont {Y.}~\bibnamefont {Wang}}, \ and\ \bibinfo
  {author} {\bibfnamefont {X.}~\bibnamefont {Zhang}},\ }\href {\doibase
  10.1126/science.aar2711} {\bibfield  {journal} {\bibinfo  {journal}
  {Science}\ }\textbf {\bibinfo {volume} {359}},\ \bibinfo {pages} {579}
  (\bibinfo {year} {2018})}\BibitemShut {NoStop}%
\bibitem [{\citenamefont {Miao}\ \emph {et~al.}(2018)\citenamefont {Miao},
  \citenamefont {Zhang}, \citenamefont {Wang}, \citenamefont {Meyers},
  \citenamefont {Said}, \citenamefont {Wang}, \citenamefont {Shi},
  \citenamefont {Weng}, \citenamefont {Fang},\ and\ \citenamefont
  {Dean}}]{Miao2018}%
  \BibitemOpen
  \bibfield  {author} {\bibinfo {author} {\bibfnamefont {H.}~\bibnamefont
  {Miao}}, \bibinfo {author} {\bibfnamefont {T.~T.}\ \bibnamefont {Zhang}},
  \bibinfo {author} {\bibfnamefont {L.}~\bibnamefont {Wang}}, \bibinfo {author}
  {\bibfnamefont {D.}~\bibnamefont {Meyers}}, \bibinfo {author} {\bibfnamefont
  {A.~H.}\ \bibnamefont {Said}}, \bibinfo {author} {\bibfnamefont {Y.~L.}\
  \bibnamefont {Wang}}, \bibinfo {author} {\bibfnamefont {Y.~G.}\ \bibnamefont
  {Shi}}, \bibinfo {author} {\bibfnamefont {H.~M.}\ \bibnamefont {Weng}},
  \bibinfo {author} {\bibfnamefont {Z.}~\bibnamefont {Fang}}, \ and\ \bibinfo
  {author} {\bibfnamefont {M.~P.~M.}\ \bibnamefont {Dean}},\ }\href {\doibase
  10.1103/PhysRevLett.121.035302} {\bibfield  {journal} {\bibinfo  {journal}
  {Phys. Rev. Lett.}\ }\textbf {\bibinfo {volume} {121}},\ \bibinfo {pages}
  {035302} (\bibinfo {year} {2018})}\BibitemShut {NoStop}%
\bibitem [{\citenamefont {Li}\ \emph {et~al.}(2021)\citenamefont {Li},
  \citenamefont {Zhang}, \citenamefont {Said}, \citenamefont {Fu},
  \citenamefont {Fabbris}, \citenamefont {Mazzone}, \citenamefont {Zhang},
  \citenamefont {Lapano}, \citenamefont {Lee}, \citenamefont {Lei},
  \citenamefont {Dean}, \citenamefont {Murakami},\ and\ \citenamefont
  {Miao}}]{Li2021}%
  \BibitemOpen
  \bibfield  {author} {\bibinfo {author} {\bibfnamefont {H.}~\bibnamefont
  {Li}}, \bibinfo {author} {\bibfnamefont {T.}~\bibnamefont {Zhang}}, \bibinfo
  {author} {\bibfnamefont {A.}~\bibnamefont {Said}}, \bibinfo {author}
  {\bibfnamefont {Y.}~\bibnamefont {Fu}}, \bibinfo {author} {\bibfnamefont
  {G.}~\bibnamefont {Fabbris}}, \bibinfo {author} {\bibfnamefont {D.~G.}\
  \bibnamefont {Mazzone}}, \bibinfo {author} {\bibfnamefont {J.}~\bibnamefont
  {Zhang}}, \bibinfo {author} {\bibfnamefont {J.}~\bibnamefont {Lapano}},
  \bibinfo {author} {\bibfnamefont {H.~N.}\ \bibnamefont {Lee}}, \bibinfo
  {author} {\bibfnamefont {H.~C.}\ \bibnamefont {Lei}}, \bibinfo {author}
  {\bibfnamefont {M.~P.~M.}\ \bibnamefont {Dean}}, \bibinfo {author}
  {\bibfnamefont {S.}~\bibnamefont {Murakami}}, \ and\ \bibinfo {author}
  {\bibfnamefont {H.}~\bibnamefont {Miao}},\ }\href {\doibase
  10.1103/PhysRevB.103.184301} {\bibfield  {journal} {\bibinfo  {journal}
  {Phys. Rev. B}\ }\textbf {\bibinfo {volume} {103}},\ \bibinfo {pages}
  {184301} (\bibinfo {year} {2021})}\BibitemShut {NoStop}%
\bibitem [{\citenamefont {Cheng}\ \emph {et~al.}(2020)\citenamefont {Cheng},
  \citenamefont {Schumann}, \citenamefont {Wang}, \citenamefont {Zhang},
  \citenamefont {Barbalas}, \citenamefont {Stemmer},\ and\ \citenamefont
  {Armitage}}]{Cheng2020}%
  \BibitemOpen
  \bibfield  {author} {\bibinfo {author} {\bibfnamefont {B.}~\bibnamefont
  {Cheng}}, \bibinfo {author} {\bibfnamefont {T.}~\bibnamefont {Schumann}},
  \bibinfo {author} {\bibfnamefont {Y.}~\bibnamefont {Wang}}, \bibinfo {author}
  {\bibfnamefont {X.}~\bibnamefont {Zhang}}, \bibinfo {author} {\bibfnamefont
  {D.}~\bibnamefont {Barbalas}}, \bibinfo {author} {\bibfnamefont
  {S.}~\bibnamefont {Stemmer}}, \ and\ \bibinfo {author} {\bibfnamefont
  {N.~P.}\ \bibnamefont {Armitage}},\ }\href {\doibase
  10.1021/acs.nanolett.0c01983} {\bibfield  {journal} {\bibinfo  {journal}
  {Nano Lett.}\ }\textbf {\bibinfo {volume} {20}},\ \bibinfo {pages} {5991}
  (\bibinfo {year} {2020})}\BibitemShut {NoStop}%
\bibitem [{\citenamefont {Lujan}\ \emph {et~al.}(2024)\citenamefont {Lujan},
  \citenamefont {Choe}, \citenamefont {Chaudhary}, \citenamefont {Ye},
  \citenamefont {Nnokwe}, \citenamefont {Rodriguez-Vega}, \citenamefont {He},
  \citenamefont {Gao}, \citenamefont {Nunley}, \citenamefont {Baldini},
  \citenamefont {Zhou}, \citenamefont {Fiete}, \citenamefont {He},\ and\
  \citenamefont {Li}}]{Lujan2024}%
  \BibitemOpen
  \bibfield  {author} {\bibinfo {author} {\bibfnamefont {D.}~\bibnamefont
  {Lujan}}, \bibinfo {author} {\bibfnamefont {J.}~\bibnamefont {Choe}},
  \bibinfo {author} {\bibfnamefont {S.}~\bibnamefont {Chaudhary}}, \bibinfo
  {author} {\bibfnamefont {G.}~\bibnamefont {Ye}}, \bibinfo {author}
  {\bibfnamefont {C.}~\bibnamefont {Nnokwe}}, \bibinfo {author} {\bibfnamefont
  {M.}~\bibnamefont {Rodriguez-Vega}}, \bibinfo {author} {\bibfnamefont
  {J.}~\bibnamefont {He}}, \bibinfo {author} {\bibfnamefont {F.~Y.}\
  \bibnamefont {Gao}}, \bibinfo {author} {\bibfnamefont {T.~N.}\ \bibnamefont
  {Nunley}}, \bibinfo {author} {\bibfnamefont {E.}~\bibnamefont {Baldini}},
  \bibinfo {author} {\bibfnamefont {J.}~\bibnamefont {Zhou}}, \bibinfo {author}
  {\bibfnamefont {G.~A.}\ \bibnamefont {Fiete}}, \bibinfo {author}
  {\bibfnamefont {R.}~\bibnamefont {He}}, \ and\ \bibinfo {author}
  {\bibfnamefont {X.}~\bibnamefont {Li}},\ }\href {\doibase
  10.1073/pnas.2304360121} {\bibfield  {journal} {\bibinfo  {journal} {Proc.
  Natl. Acad. Sci.}\ }\textbf {\bibinfo {volume} {121}} (\bibinfo {year}
  {2024}),\ 10.1073/pnas.2304360121}\BibitemShut {NoStop}%
\bibitem [{\citenamefont {Hernandez}\ \emph {et~al.}(2023)\citenamefont
  {Hernandez}, \citenamefont {Baydin}, \citenamefont {Chaudhary}, \citenamefont
  {Tay}, \citenamefont {Katayama}, \citenamefont {Takeda}, \citenamefont
  {Nojiri}, \citenamefont {Okazaki}, \citenamefont {Rappl}, \citenamefont
  {Abramof}, \citenamefont {Rodriguez-Vega}, \citenamefont {Fiete},\ and\
  \citenamefont {Kono}}]{Hernandez2023}%
  \BibitemOpen
  \bibfield  {author} {\bibinfo {author} {\bibfnamefont {F.~G.~G.}\
  \bibnamefont {Hernandez}}, \bibinfo {author} {\bibfnamefont {A.}~\bibnamefont
  {Baydin}}, \bibinfo {author} {\bibfnamefont {S.}~\bibnamefont {Chaudhary}},
  \bibinfo {author} {\bibfnamefont {F.}~\bibnamefont {Tay}}, \bibinfo {author}
  {\bibfnamefont {I.}~\bibnamefont {Katayama}}, \bibinfo {author}
  {\bibfnamefont {J.}~\bibnamefont {Takeda}}, \bibinfo {author} {\bibfnamefont
  {H.}~\bibnamefont {Nojiri}}, \bibinfo {author} {\bibfnamefont {A.~K.}\
  \bibnamefont {Okazaki}}, \bibinfo {author} {\bibfnamefont {P.~H.~O.}\
  \bibnamefont {Rappl}}, \bibinfo {author} {\bibfnamefont {E.}~\bibnamefont
  {Abramof}}, \bibinfo {author} {\bibfnamefont {M.}~\bibnamefont
  {Rodriguez-Vega}}, \bibinfo {author} {\bibfnamefont {G.~A.}\ \bibnamefont
  {Fiete}}, \ and\ \bibinfo {author} {\bibfnamefont {J.}~\bibnamefont {Kono}},\
  }\href {\doibase 10.1126/sciadv.adj4074} {\bibfield  {journal} {\bibinfo
  {journal} {Sci. Adv.}\ }\textbf {\bibinfo {volume} {9}} (\bibinfo {year}
  {2023}),\ 10.1126/sciadv.adj4074}\BibitemShut {NoStop}%
\bibitem [{\citenamefont {Wu}\ \emph {et~al.}(2023)\citenamefont {Wu},
  \citenamefont {Bao}, \citenamefont {Zhou}, \citenamefont {Wang},
  \citenamefont {Sun}, \citenamefont {Wen}, \citenamefont {Wan},\ and\
  \citenamefont {Zhang}}]{Wu2023}%
  \BibitemOpen
  \bibfield  {author} {\bibinfo {author} {\bibfnamefont {F.}~\bibnamefont
  {Wu}}, \bibinfo {author} {\bibfnamefont {S.}~\bibnamefont {Bao}}, \bibinfo
  {author} {\bibfnamefont {J.}~\bibnamefont {Zhou}}, \bibinfo {author}
  {\bibfnamefont {Y.}~\bibnamefont {Wang}}, \bibinfo {author} {\bibfnamefont
  {J.}~\bibnamefont {Sun}}, \bibinfo {author} {\bibfnamefont {J.}~\bibnamefont
  {Wen}}, \bibinfo {author} {\bibfnamefont {Y.}~\bibnamefont {Wan}}, \ and\
  \bibinfo {author} {\bibfnamefont {Q.}~\bibnamefont {Zhang}},\ }\href
  {\doibase 10.1038/s41567-023-02210-4} {\bibfield  {journal} {\bibinfo
  {journal} {Nat. Phys.}\ } (\bibinfo {year} {2023}),\
  10.1038/s41567-023-02210-4}\BibitemShut {NoStop}%
\bibitem [{\citenamefont {Schaack}(1976)}]{Schaack1976}%
  \BibitemOpen
  \bibfield  {author} {\bibinfo {author} {\bibfnamefont {G.}~\bibnamefont
  {Schaack}},\ }\href {\doibase 10.1088/0022-3719/9/11/009} {\bibfield
  {journal} {\bibinfo  {journal} {J. Phys. C Solid State Phys.}\ }\textbf
  {\bibinfo {volume} {9}},\ \bibinfo {pages} {L297} (\bibinfo {year}
  {1976})}\BibitemShut {NoStop}%
\bibitem [{\citenamefont {Luo}\ \emph {et~al.}(2023)\citenamefont {Luo},
  \citenamefont {Lin}, \citenamefont {Zhang}, \citenamefont {Chen},
  \citenamefont {Blackert}, \citenamefont {Xu}, \citenamefont {Yakobson},\ and\
  \citenamefont {Zhu}}]{Luo2023}%
  \BibitemOpen
  \bibfield  {author} {\bibinfo {author} {\bibfnamefont {J.}~\bibnamefont
  {Luo}}, \bibinfo {author} {\bibfnamefont {T.}~\bibnamefont {Lin}}, \bibinfo
  {author} {\bibfnamefont {J.}~\bibnamefont {Zhang}}, \bibinfo {author}
  {\bibfnamefont {X.}~\bibnamefont {Chen}}, \bibinfo {author} {\bibfnamefont
  {E.~R.}\ \bibnamefont {Blackert}}, \bibinfo {author} {\bibfnamefont
  {R.}~\bibnamefont {Xu}}, \bibinfo {author} {\bibfnamefont {B.~I.}\
  \bibnamefont {Yakobson}}, \ and\ \bibinfo {author} {\bibfnamefont
  {H.}~\bibnamefont {Zhu}},\ }\href {\doibase 10.1126/science.adi9601}
  {\bibfield  {journal} {\bibinfo  {journal} {Science}\ }\textbf {\bibinfo
  {volume} {382}},\ \bibinfo {pages} {698} (\bibinfo {year}
  {2023})}\BibitemShut {NoStop}%
\bibitem [{\citenamefont {Baydin}\ \emph {et~al.}(2022)\citenamefont {Baydin},
  \citenamefont {Hernandez}, \citenamefont {Rodriguez-Vega}, \citenamefont
  {Okazaki}, \citenamefont {Tay}, \citenamefont {Noe}, \citenamefont
  {Katayama}, \citenamefont {Takeda}, \citenamefont {Nojiri}, \citenamefont
  {Rappl}, \citenamefont {Abramof}, \citenamefont {Fiete},\ and\ \citenamefont
  {Kono}}]{Baydin2022}%
  \BibitemOpen
  \bibfield  {author} {\bibinfo {author} {\bibfnamefont {A.}~\bibnamefont
  {Baydin}}, \bibinfo {author} {\bibfnamefont {F.~G.~G.}\ \bibnamefont
  {Hernandez}}, \bibinfo {author} {\bibfnamefont {M.}~\bibnamefont
  {Rodriguez-Vega}}, \bibinfo {author} {\bibfnamefont {A.~K.}\ \bibnamefont
  {Okazaki}}, \bibinfo {author} {\bibfnamefont {F.}~\bibnamefont {Tay}},
  \bibinfo {author} {\bibfnamefont {G.~T.}\ \bibnamefont {Noe}}, \bibinfo
  {author} {\bibfnamefont {I.}~\bibnamefont {Katayama}}, \bibinfo {author}
  {\bibfnamefont {J.}~\bibnamefont {Takeda}}, \bibinfo {author} {\bibfnamefont
  {H.}~\bibnamefont {Nojiri}}, \bibinfo {author} {\bibfnamefont {P.~H.~O.}\
  \bibnamefont {Rappl}}, \bibinfo {author} {\bibfnamefont {E.}~\bibnamefont
  {Abramof}}, \bibinfo {author} {\bibfnamefont {G.~A.}\ \bibnamefont {Fiete}},
  \ and\ \bibinfo {author} {\bibfnamefont {J.}~\bibnamefont {Kono}},\ }\href
  {\doibase 10.1103/PhysRevLett.128.075901} {\bibfield  {journal} {\bibinfo
  {journal} {Phys. Rev. Lett.}\ }\textbf {\bibinfo {volume} {128}},\ \bibinfo
  {pages} {075901} (\bibinfo {year} {2022})}\BibitemShut {NoStop}%
\bibitem [{\citenamefont {Liu}\ and\ \citenamefont {Shi}(2017)}]{Liu2017}%
  \BibitemOpen
  \bibfield  {author} {\bibinfo {author} {\bibfnamefont {D.}~\bibnamefont
  {Liu}}\ and\ \bibinfo {author} {\bibfnamefont {J.}~\bibnamefont {Shi}},\
  }\href {\doibase 10.1103/PhysRevLett.119.075301} {\bibfield  {journal}
  {\bibinfo  {journal} {Phys. Rev. Lett.}\ }\textbf {\bibinfo {volume} {119}},\
  \bibinfo {pages} {075301} (\bibinfo {year} {2017})}\BibitemShut {NoStop}%
\bibitem [{\citenamefont {Bistoni}\ \emph {et~al.}(2021)\citenamefont
  {Bistoni}, \citenamefont {Mauri},\ and\ \citenamefont
  {Calandra}}]{Bistoni2021}%
  \BibitemOpen
  \bibfield  {author} {\bibinfo {author} {\bibfnamefont {O.}~\bibnamefont
  {Bistoni}}, \bibinfo {author} {\bibfnamefont {F.}~\bibnamefont {Mauri}}, \
  and\ \bibinfo {author} {\bibfnamefont {M.}~\bibnamefont {Calandra}},\ }\href
  {\doibase 10.1103/PhysRevLett.126.225703} {\bibfield  {journal} {\bibinfo
  {journal} {Phys. Rev. Lett.}\ }\textbf {\bibinfo {volume} {126}},\ \bibinfo
  {pages} {225703} (\bibinfo {year} {2021})}\BibitemShut {NoStop}%
\bibitem [{\citenamefont {Yang}\ \emph {et~al.}(2017)\citenamefont {Yang},
  \citenamefont {Dai}, \citenamefont {Xu}, \citenamefont {Zhang}, \citenamefont
  {Qiu}, \citenamefont {Sui}, \citenamefont {Homes},\ and\ \citenamefont
  {Qiu}}]{Yang2017}%
  \BibitemOpen
  \bibfield  {author} {\bibinfo {author} {\bibfnamefont {R.}~\bibnamefont
  {Yang}}, \bibinfo {author} {\bibfnamefont {Y.}~\bibnamefont {Dai}}, \bibinfo
  {author} {\bibfnamefont {B.}~\bibnamefont {Xu}}, \bibinfo {author}
  {\bibfnamefont {W.}~\bibnamefont {Zhang}}, \bibinfo {author} {\bibfnamefont
  {Z.}~\bibnamefont {Qiu}}, \bibinfo {author} {\bibfnamefont {Q.}~\bibnamefont
  {Sui}}, \bibinfo {author} {\bibfnamefont {C.~C.}\ \bibnamefont {Homes}}, \
  and\ \bibinfo {author} {\bibfnamefont {X.}~\bibnamefont {Qiu}},\ }\href
  {\doibase 10.1103/PhysRevB.95.064506} {\bibfield  {journal} {\bibinfo
  {journal} {Phys. Rev. B}\ }\textbf {\bibinfo {volume} {95}},\ \bibinfo
  {pages} {064506} (\bibinfo {year} {2017})}\BibitemShut {NoStop}%
\bibitem [{\citenamefont {Xu}\ \emph {et~al.}(2017)\citenamefont {Xu},
  \citenamefont {Dai}, \citenamefont {Zhao}, \citenamefont {Wang},
  \citenamefont {Yang}, \citenamefont {Zhang}, \citenamefont {Liu},
  \citenamefont {Xiao}, \citenamefont {Chen}, \citenamefont {Trugman},
  \citenamefont {Zhu}, \citenamefont {Taylor}, \citenamefont {Yarotski},
  \citenamefont {Prasankumar},\ and\ \citenamefont {Qiu}}]{Xu2017}%
  \BibitemOpen
  \bibfield  {author} {\bibinfo {author} {\bibfnamefont {B.}~\bibnamefont
  {Xu}}, \bibinfo {author} {\bibfnamefont {Y.~M.}\ \bibnamefont {Dai}},
  \bibinfo {author} {\bibfnamefont {L.~X.}\ \bibnamefont {Zhao}}, \bibinfo
  {author} {\bibfnamefont {K.}~\bibnamefont {Wang}}, \bibinfo {author}
  {\bibfnamefont {R.}~\bibnamefont {Yang}}, \bibinfo {author} {\bibfnamefont
  {W.}~\bibnamefont {Zhang}}, \bibinfo {author} {\bibfnamefont {J.~Y.}\
  \bibnamefont {Liu}}, \bibinfo {author} {\bibfnamefont {H.}~\bibnamefont
  {Xiao}}, \bibinfo {author} {\bibfnamefont {G.~F.}\ \bibnamefont {Chen}},
  \bibinfo {author} {\bibfnamefont {S.~A.}\ \bibnamefont {Trugman}}, \bibinfo
  {author} {\bibfnamefont {J.~X.}\ \bibnamefont {Zhu}}, \bibinfo {author}
  {\bibfnamefont {A.~J.}\ \bibnamefont {Taylor}}, \bibinfo {author}
  {\bibfnamefont {D.~A.}\ \bibnamefont {Yarotski}}, \bibinfo {author}
  {\bibfnamefont {R.~P.}\ \bibnamefont {Prasankumar}}, \ and\ \bibinfo {author}
  {\bibfnamefont {X.~G.}\ \bibnamefont {Qiu}},\ }\href {\doibase
  10.1038/ncomms14933} {\bibfield  {journal} {\bibinfo  {journal} {Nat.
  Commun.}\ }\textbf {\bibinfo {volume} {8}},\ \bibinfo {pages} {14933}
  (\bibinfo {year} {2017})}\BibitemShut {NoStop}%
\bibitem [{\citenamefont {Cappelluti}\ \emph {et~al.}(2012)\citenamefont
  {Cappelluti}, \citenamefont {Benfatto}, \citenamefont {Manzardo},\ and\
  \citenamefont {Kuzmenko}}]{Cappelluti2012}%
  \BibitemOpen
  \bibfield  {author} {\bibinfo {author} {\bibfnamefont {E.}~\bibnamefont
  {Cappelluti}}, \bibinfo {author} {\bibfnamefont {L.}~\bibnamefont
  {Benfatto}}, \bibinfo {author} {\bibfnamefont {M.}~\bibnamefont {Manzardo}},
  \ and\ \bibinfo {author} {\bibfnamefont {A.~B.}\ \bibnamefont {Kuzmenko}},\
  }\href {\doibase 10.1103/PhysRevB.86.115439} {\bibfield  {journal} {\bibinfo
  {journal} {Phys. Rev. B}\ }\textbf {\bibinfo {volume} {86}},\ \bibinfo
  {pages} {115439} (\bibinfo {year} {2012})}\BibitemShut {NoStop}%
\bibitem [{\citenamefont {Okamura}\ \emph {et~al.}(2020)\citenamefont
  {Okamura}, \citenamefont {Minami}, \citenamefont {Kato}, \citenamefont
  {Fujishiro}, \citenamefont {Kaneko}, \citenamefont {Ikeda}, \citenamefont
  {Muramoto}, \citenamefont {Kaneko}, \citenamefont {Ueda}, \citenamefont
  {Kocsis}, \citenamefont {Kanazawa}, \citenamefont {Taguchi}, \citenamefont
  {Koretsune}, \citenamefont {Fujiwara}, \citenamefont {Tsukazaki},
  \citenamefont {Arita}, \citenamefont {Tokura},\ and\ \citenamefont
  {Takahashi}}]{Okamura2020}%
  \BibitemOpen
  \bibfield  {author} {\bibinfo {author} {\bibfnamefont {Y.}~\bibnamefont
  {Okamura}}, \bibinfo {author} {\bibfnamefont {S.}~\bibnamefont {Minami}},
  \bibinfo {author} {\bibfnamefont {Y.}~\bibnamefont {Kato}}, \bibinfo {author}
  {\bibfnamefont {Y.}~\bibnamefont {Fujishiro}}, \bibinfo {author}
  {\bibfnamefont {Y.}~\bibnamefont {Kaneko}}, \bibinfo {author} {\bibfnamefont
  {J.}~\bibnamefont {Ikeda}}, \bibinfo {author} {\bibfnamefont
  {J.}~\bibnamefont {Muramoto}}, \bibinfo {author} {\bibfnamefont
  {R.}~\bibnamefont {Kaneko}}, \bibinfo {author} {\bibfnamefont
  {K.}~\bibnamefont {Ueda}}, \bibinfo {author} {\bibfnamefont {V.}~\bibnamefont
  {Kocsis}}, \bibinfo {author} {\bibfnamefont {N.}~\bibnamefont {Kanazawa}},
  \bibinfo {author} {\bibfnamefont {Y.}~\bibnamefont {Taguchi}}, \bibinfo
  {author} {\bibfnamefont {T.}~\bibnamefont {Koretsune}}, \bibinfo {author}
  {\bibfnamefont {K.}~\bibnamefont {Fujiwara}}, \bibinfo {author}
  {\bibfnamefont {A.}~\bibnamefont {Tsukazaki}}, \bibinfo {author}
  {\bibfnamefont {R.}~\bibnamefont {Arita}}, \bibinfo {author} {\bibfnamefont
  {Y.}~\bibnamefont {Tokura}}, \ and\ \bibinfo {author} {\bibfnamefont
  {Y.}~\bibnamefont {Takahashi}},\ }\href {\doibase 10.1038/s41467-020-18470-0}
  {\bibfield  {journal} {\bibinfo  {journal} {Nat. Commun.}\ }\textbf {\bibinfo
  {volume} {11}},\ \bibinfo {pages} {4619} (\bibinfo {year}
  {2020})}\BibitemShut {NoStop}%
\bibitem [{\citenamefont {Levallois}\ \emph {et~al.}(2015)\citenamefont
  {Levallois}, \citenamefont {Nedoliuk}, \citenamefont {Crassee},\ and\
  \citenamefont {Kuzmenko}}]{Levallois2015}%
  \BibitemOpen
  \bibfield  {author} {\bibinfo {author} {\bibfnamefont {J.}~\bibnamefont
  {Levallois}}, \bibinfo {author} {\bibfnamefont {I.~O.}\ \bibnamefont
  {Nedoliuk}}, \bibinfo {author} {\bibfnamefont {I.}~\bibnamefont {Crassee}}, \
  and\ \bibinfo {author} {\bibfnamefont {A.~B.}\ \bibnamefont {Kuzmenko}},\
  }\href {\doibase 10.1063/1.4914846} {\bibfield  {journal} {\bibinfo
  {journal} {Rev. of Sci. Instrum.}\ }\textbf {\bibinfo {volume} {86}}
  (\bibinfo {year} {2015}),\ 10.1063/1.4914846}\BibitemShut {NoStop}%
\bibitem [{Note1()}]{Note1}%
  \BibitemOpen
  \bibinfo {note} {See Supplemental Material at
  http://link.aps.org/supplemental for the sample characterizations and
  experimental techniques as well as for complementary data and analysis, which
  include Refs.~\cite {Yang2020, Okamura2020, Levallois2015, Soh2022, Li2012,
  Cappelluti2012, Ren2021, Saparov2022}}\BibitemShut {NoStop}%
\bibitem [{\citenamefont {Yang}\ \emph {et~al.}(2020)\citenamefont {Yang},
  \citenamefont {Zhang}, \citenamefont {Zhou}, \citenamefont {Dai},
  \citenamefont {Liao}, \citenamefont {Weng},\ and\ \citenamefont
  {Qiu}}]{Yang2020}%
  \BibitemOpen
  \bibfield  {author} {\bibinfo {author} {\bibfnamefont {R.}~\bibnamefont
  {Yang}}, \bibinfo {author} {\bibfnamefont {T.}~\bibnamefont {Zhang}},
  \bibinfo {author} {\bibfnamefont {L.}~\bibnamefont {Zhou}}, \bibinfo {author}
  {\bibfnamefont {Y.}~\bibnamefont {Dai}}, \bibinfo {author} {\bibfnamefont
  {Z.}~\bibnamefont {Liao}}, \bibinfo {author} {\bibfnamefont {H.}~\bibnamefont
  {Weng}}, \ and\ \bibinfo {author} {\bibfnamefont {X.}~\bibnamefont {Qiu}},\
  }\href {\doibase 10.1103/PhysRevLett.124.077403} {\bibfield  {journal}
  {\bibinfo  {journal} {Phys. Rev. Lett.}\ }\textbf {\bibinfo {volume} {124}},\
  \bibinfo {pages} {077403} (\bibinfo {year} {2020})}\BibitemShut {NoStop}%
\bibitem [{\citenamefont {Xu}\ \emph {et~al.}(2020)\citenamefont {Xu},
  \citenamefont {Zhao}, \citenamefont {Yi}, \citenamefont {Wang}, \citenamefont
  {Yin}, \citenamefont {Wang}, \citenamefont {Hu}, \citenamefont {Wang},
  \citenamefont {Liu}, \citenamefont {Xu}, \citenamefont {Lu}, \citenamefont
  {Soluyanov}, \citenamefont {Lei}, \citenamefont {Shi}, \citenamefont {Luo},\
  and\ \citenamefont {Chen}}]{XuYS2020}%
  \BibitemOpen
  \bibfield  {author} {\bibinfo {author} {\bibfnamefont {Y.}~\bibnamefont
  {Xu}}, \bibinfo {author} {\bibfnamefont {J.}~\bibnamefont {Zhao}}, \bibinfo
  {author} {\bibfnamefont {C.}~\bibnamefont {Yi}}, \bibinfo {author}
  {\bibfnamefont {Q.}~\bibnamefont {Wang}}, \bibinfo {author} {\bibfnamefont
  {Q.}~\bibnamefont {Yin}}, \bibinfo {author} {\bibfnamefont {Y.}~\bibnamefont
  {Wang}}, \bibinfo {author} {\bibfnamefont {X.}~\bibnamefont {Hu}}, \bibinfo
  {author} {\bibfnamefont {L.}~\bibnamefont {Wang}}, \bibinfo {author}
  {\bibfnamefont {E.}~\bibnamefont {Liu}}, \bibinfo {author} {\bibfnamefont
  {G.}~\bibnamefont {Xu}}, \bibinfo {author} {\bibfnamefont {L.}~\bibnamefont
  {Lu}}, \bibinfo {author} {\bibfnamefont {A.~A.}\ \bibnamefont {Soluyanov}},
  \bibinfo {author} {\bibfnamefont {H.}~\bibnamefont {Lei}}, \bibinfo {author}
  {\bibfnamefont {Y.}~\bibnamefont {Shi}}, \bibinfo {author} {\bibfnamefont
  {J.}~\bibnamefont {Luo}}, \ and\ \bibinfo {author} {\bibfnamefont {Z.~G.}\
  \bibnamefont {Chen}},\ }\href {\doibase 10.1038/s41467-020-17234-0}
  {\bibfield  {journal} {\bibinfo  {journal} {Nat Commun}\ }\textbf {\bibinfo
  {volume} {11}},\ \bibinfo {pages} {3985} (\bibinfo {year}
  {2020})}\BibitemShut {NoStop}%
\bibitem [{\citenamefont {Li}\ \emph {et~al.}(2012)\citenamefont {Li},
  \citenamefont {Lui}, \citenamefont {Cappelluti}, \citenamefont {Benfatto},
  \citenamefont {Mak}, \citenamefont {Carr}, \citenamefont {Shan},\ and\
  \citenamefont {Heinz}}]{Li2012}%
  \BibitemOpen
  \bibfield  {author} {\bibinfo {author} {\bibfnamefont {Z.}~\bibnamefont
  {Li}}, \bibinfo {author} {\bibfnamefont {C.~H.}\ \bibnamefont {Lui}},
  \bibinfo {author} {\bibfnamefont {E.}~\bibnamefont {Cappelluti}}, \bibinfo
  {author} {\bibfnamefont {L.}~\bibnamefont {Benfatto}}, \bibinfo {author}
  {\bibfnamefont {K.~F.}\ \bibnamefont {Mak}}, \bibinfo {author} {\bibfnamefont
  {G.~L.}\ \bibnamefont {Carr}}, \bibinfo {author} {\bibfnamefont
  {J.}~\bibnamefont {Shan}}, \ and\ \bibinfo {author} {\bibfnamefont {T.~F.}\
  \bibnamefont {Heinz}},\ }\href {\doibase 10.1103/PhysRevLett.108.156801}
  {\bibfield  {journal} {\bibinfo  {journal} {Phys. Rev. Lett.}\ }\textbf
  {\bibinfo {volume} {108}},\ \bibinfo {pages} {156801} (\bibinfo {year}
  {2012})}\BibitemShut {NoStop}%
\bibitem [{\citenamefont {Xu}\ \emph {et~al.}(2019)\citenamefont {Xu},
  \citenamefont {Cappelluti}, \citenamefont {Benfatto}, \citenamefont
  {Mallett}, \citenamefont {Marsik}, \citenamefont {Sheveleva}, \citenamefont
  {Lyzwa}, \citenamefont {Wolf}, \citenamefont {Yang}, \citenamefont {Qiu},
  \citenamefont {Dai}, \citenamefont {Wen}, \citenamefont {Lobo},\ and\
  \citenamefont {Bernhard}}]{Xu2019}%
  \BibitemOpen
  \bibfield  {author} {\bibinfo {author} {\bibfnamefont {B.}~\bibnamefont
  {Xu}}, \bibinfo {author} {\bibfnamefont {E.}~\bibnamefont {Cappelluti}},
  \bibinfo {author} {\bibfnamefont {L.}~\bibnamefont {Benfatto}}, \bibinfo
  {author} {\bibfnamefont {B.~P.~P.}\ \bibnamefont {Mallett}}, \bibinfo
  {author} {\bibfnamefont {P.}~\bibnamefont {Marsik}}, \bibinfo {author}
  {\bibfnamefont {E.}~\bibnamefont {Sheveleva}}, \bibinfo {author}
  {\bibfnamefont {F.}~\bibnamefont {Lyzwa}}, \bibinfo {author} {\bibfnamefont
  {T.}~\bibnamefont {Wolf}}, \bibinfo {author} {\bibfnamefont {R.}~\bibnamefont
  {Yang}}, \bibinfo {author} {\bibfnamefont {X.~G.}\ \bibnamefont {Qiu}},
  \bibinfo {author} {\bibfnamefont {Y.~M.}\ \bibnamefont {Dai}}, \bibinfo
  {author} {\bibfnamefont {H.~H.}\ \bibnamefont {Wen}}, \bibinfo {author}
  {\bibfnamefont {R.~P. S.~M.}\ \bibnamefont {Lobo}}, \ and\ \bibinfo {author}
  {\bibfnamefont {C.}~\bibnamefont {Bernhard}},\ }\href {\doibase
  10.1103/PhysRevLett.122.217002} {\bibfield  {journal} {\bibinfo  {journal}
  {Phy. Rev. Lett.}\ }\textbf {\bibinfo {volume} {122}},\ \bibinfo {pages}
  {217002} (\bibinfo {year} {2019})}\BibitemShut {NoStop}%
\bibitem [{Note2()}]{Note2}%
  \BibitemOpen
  \bibinfo {note} {Since these phonons are infrared active at the Brillouin
  zone center, they interfere with the direct interband transition as shown in
  Fig.~2d, which do not show momentum change.}\BibitemShut {Stop}%
\bibitem [{\citenamefont {Ma}\ \emph {et~al.}(2017)\citenamefont {Ma},
  \citenamefont {Xu}, \citenamefont {Chan}, \citenamefont {Zhang},
  \citenamefont {Chang}, \citenamefont {Lin}, \citenamefont {Xie},
  \citenamefont {Palacios}, \citenamefont {Lin}, \citenamefont {Jia},
  \citenamefont {Lee}, \citenamefont {Jarillo-Herrero},\ and\ \citenamefont
  {Gedik}}]{Ma2017}%
  \BibitemOpen
  \bibfield  {author} {\bibinfo {author} {\bibfnamefont {Q.}~\bibnamefont
  {Ma}}, \bibinfo {author} {\bibfnamefont {S.-Y.}\ \bibnamefont {Xu}}, \bibinfo
  {author} {\bibfnamefont {C.-K.}\ \bibnamefont {Chan}}, \bibinfo {author}
  {\bibfnamefont {C.-L.}\ \bibnamefont {Zhang}}, \bibinfo {author}
  {\bibfnamefont {G.}~\bibnamefont {Chang}}, \bibinfo {author} {\bibfnamefont
  {Y.}~\bibnamefont {Lin}}, \bibinfo {author} {\bibfnamefont {W.}~\bibnamefont
  {Xie}}, \bibinfo {author} {\bibfnamefont {T.}~\bibnamefont {Palacios}},
  \bibinfo {author} {\bibfnamefont {H.}~\bibnamefont {Lin}}, \bibinfo {author}
  {\bibfnamefont {S.}~\bibnamefont {Jia}}, \bibinfo {author} {\bibfnamefont
  {P.~A.}\ \bibnamefont {Lee}}, \bibinfo {author} {\bibfnamefont
  {P.}~\bibnamefont {Jarillo-Herrero}}, \ and\ \bibinfo {author} {\bibfnamefont
  {N.}~\bibnamefont {Gedik}},\ }\href {\doibase 10.1038/nphys4146} {\bibfield
  {journal} {\bibinfo  {journal} {Nat. Phys.}\ }\textbf {\bibinfo {volume}
  {13}},\ \bibinfo {pages} {842} (\bibinfo {year} {2017})}\BibitemShut
  {NoStop}%
\bibitem [{\citenamefont {Sugano}\ and\ \citenamefont
  {Kojima}(2000)}]{Sugano2000}%
  \BibitemOpen
  \bibinfo {editor} {\bibfnamefont {S.}~\bibnamefont {Sugano}}\ and\ \bibinfo
  {editor} {\bibfnamefont {N.}~\bibnamefont {Kojima}},\ eds.,\ \href {\doibase
  10.1007/978-3-662-04143-7} {\emph {\bibinfo {title} {{Magneto-Optics}}}},\
  \bibinfo {series} {Springer Series in Solid-State Sciences}, Vol.\ \bibinfo
  {volume} {128}\ (\bibinfo  {publisher} {Springer Berlin Heidelberg},\
  \bibinfo {address} {Berlin, Heidelberg},\ \bibinfo {year} {2000})\BibitemShut
  {NoStop}%
\bibitem [{Note3()}]{Note3}%
  \BibitemOpen
  \bibinfo {note} {Since there is no anisotropy within Co$_3$Sn$_2$S$_2$'s
  \protect \emph {ab}-plane, $\sigma _{xx1}(\omega )$ is equal to $\sigma
  _{1}(\omega )$.}\BibitemShut {Stop}%
\bibitem [{\citenamefont {Liu}\ \emph {et~al.}(2021)\citenamefont {Liu},
  \citenamefont {Xu}, \citenamefont {Liu}, \citenamefont {Shen}, \citenamefont
  {Le}, \citenamefont {Li}, \citenamefont {Pei}, \citenamefont {Liang},
  \citenamefont {Dudin}, \citenamefont {Kim}, \citenamefont {Cacho},
  \citenamefont {Xu}, \citenamefont {Sun}, \citenamefont {Yang}, \citenamefont
  {Liu}, \citenamefont {Felser}, \citenamefont {Parkin},\ and\ \citenamefont
  {Chen}}]{Liu2021}%
  \BibitemOpen
  \bibfield  {author} {\bibinfo {author} {\bibfnamefont {D.~F.}\ \bibnamefont
  {Liu}}, \bibinfo {author} {\bibfnamefont {Q.~N.}\ \bibnamefont {Xu}},
  \bibinfo {author} {\bibfnamefont {E.~K.}\ \bibnamefont {Liu}}, \bibinfo
  {author} {\bibfnamefont {J.~L.}\ \bibnamefont {Shen}}, \bibinfo {author}
  {\bibfnamefont {C.~C.}\ \bibnamefont {Le}}, \bibinfo {author} {\bibfnamefont
  {Y.~W.}\ \bibnamefont {Li}}, \bibinfo {author} {\bibfnamefont
  {D.}~\bibnamefont {Pei}}, \bibinfo {author} {\bibfnamefont {A.~J.}\
  \bibnamefont {Liang}}, \bibinfo {author} {\bibfnamefont {P.}~\bibnamefont
  {Dudin}}, \bibinfo {author} {\bibfnamefont {T.~K.}\ \bibnamefont {Kim}},
  \bibinfo {author} {\bibfnamefont {C.}~\bibnamefont {Cacho}}, \bibinfo
  {author} {\bibfnamefont {Y.~F.}\ \bibnamefont {Xu}}, \bibinfo {author}
  {\bibfnamefont {Y.}~\bibnamefont {Sun}}, \bibinfo {author} {\bibfnamefont
  {L.~X.}\ \bibnamefont {Yang}}, \bibinfo {author} {\bibfnamefont {Z.~K.}\
  \bibnamefont {Liu}}, \bibinfo {author} {\bibfnamefont {C.}~\bibnamefont
  {Felser}}, \bibinfo {author} {\bibfnamefont {S.~S.~P.}\ \bibnamefont
  {Parkin}}, \ and\ \bibinfo {author} {\bibfnamefont {Y.~L.}\ \bibnamefont
  {Chen}},\ }\href {\doibase 10.1103/PhysRevB.104.205140} {\bibfield  {journal}
  {\bibinfo  {journal} {Phys. Rev. B}\ }\textbf {\bibinfo {volume} {104}},\
  \bibinfo {pages} {205140} (\bibinfo {year} {2021})}\BibitemShut {NoStop}%
\bibitem [{\citenamefont {Neubauer}\ \emph {et~al.}(2022)\citenamefont
  {Neubauer}, \citenamefont {Ye}, \citenamefont {Shi}, \citenamefont
  {Malinowski}, \citenamefont {Gao}, \citenamefont {Taddei}, \citenamefont
  {Bourges}, \citenamefont {Ivanov}, \citenamefont {Chu},\ and\ \citenamefont
  {Dai}}]{Neubauer2022}%
  \BibitemOpen
  \bibfield  {author} {\bibinfo {author} {\bibfnamefont {K.~J.}\ \bibnamefont
  {Neubauer}}, \bibinfo {author} {\bibfnamefont {F.}~\bibnamefont {Ye}},
  \bibinfo {author} {\bibfnamefont {Y.}~\bibnamefont {Shi}}, \bibinfo {author}
  {\bibfnamefont {P.}~\bibnamefont {Malinowski}}, \bibinfo {author}
  {\bibfnamefont {B.}~\bibnamefont {Gao}}, \bibinfo {author} {\bibfnamefont
  {K.~M.}\ \bibnamefont {Taddei}}, \bibinfo {author} {\bibfnamefont
  {P.}~\bibnamefont {Bourges}}, \bibinfo {author} {\bibfnamefont
  {A.}~\bibnamefont {Ivanov}}, \bibinfo {author} {\bibfnamefont {J.-H.}\
  \bibnamefont {Chu}}, \ and\ \bibinfo {author} {\bibfnamefont
  {P.}~\bibnamefont {Dai}},\ }\href {\doibase 10.1038/s41535-022-00523-w}
  {\bibfield  {journal} {\bibinfo  {journal} {npj Quantum Mater.}\ }\textbf
  {\bibinfo {volume} {7}} (\bibinfo {year} {2022}),\
  10.1038/s41535-022-00523-w}\BibitemShut {NoStop}%
\bibitem [{\citenamefont {Xu}\ \emph {et~al.}(2018)\citenamefont {Xu},
  \citenamefont {Liu}, \citenamefont {Shi}, \citenamefont {Muechler},
  \citenamefont {Gayles}, \citenamefont {Felser},\ and\ \citenamefont
  {Sun}}]{XuQ2018}%
  \BibitemOpen
  \bibfield  {author} {\bibinfo {author} {\bibfnamefont {Q.}~\bibnamefont
  {Xu}}, \bibinfo {author} {\bibfnamefont {E.}~\bibnamefont {Liu}}, \bibinfo
  {author} {\bibfnamefont {W.}~\bibnamefont {Shi}}, \bibinfo {author}
  {\bibfnamefont {L.}~\bibnamefont {Muechler}}, \bibinfo {author}
  {\bibfnamefont {J.}~\bibnamefont {Gayles}}, \bibinfo {author} {\bibfnamefont
  {C.}~\bibnamefont {Felser}}, \ and\ \bibinfo {author} {\bibfnamefont
  {Y.}~\bibnamefont {Sun}},\ }\href {\doibase 10.1103/PhysRevB.97.235416}
  {\bibfield  {journal} {\bibinfo  {journal} {Phys. Rev. B}\ }\textbf {\bibinfo
  {volume} {97}} (\bibinfo {year} {2018}),\
  10.1103/PhysRevB.97.235416}\BibitemShut {NoStop}%
\bibitem [{\citenamefont {Soh}\ \emph {et~al.}(2022)\citenamefont {Soh},
  \citenamefont {Yi}, \citenamefont {Zivkovic}, \citenamefont {Qureshi},
  \citenamefont {Stunault}, \citenamefont {Ouladdiaf}, \citenamefont
  {Rodr{\'{i}}guez-Velamaz{\'{a}}n}, \citenamefont {Shi}, \citenamefont
  {R{\o}nnow},\ and\ \citenamefont {Boothroyd}}]{Soh2022}%
  \BibitemOpen
  \bibfield  {author} {\bibinfo {author} {\bibfnamefont {J.-R.}\ \bibnamefont
  {Soh}}, \bibinfo {author} {\bibfnamefont {C.}~\bibnamefont {Yi}}, \bibinfo
  {author} {\bibfnamefont {I.}~\bibnamefont {Zivkovic}}, \bibinfo {author}
  {\bibfnamefont {N.}~\bibnamefont {Qureshi}}, \bibinfo {author} {\bibfnamefont
  {A.}~\bibnamefont {Stunault}}, \bibinfo {author} {\bibfnamefont
  {B.}~\bibnamefont {Ouladdiaf}}, \bibinfo {author} {\bibfnamefont {J.~A.}\
  \bibnamefont {Rodr{\'{i}}guez-Velamaz{\'{a}}n}}, \bibinfo {author}
  {\bibfnamefont {Y.}~\bibnamefont {Shi}}, \bibinfo {author} {\bibfnamefont
  {H.~M.}\ \bibnamefont {R{\o}nnow}}, \ and\ \bibinfo {author} {\bibfnamefont
  {A.~T.}\ \bibnamefont {Boothroyd}},\ }\href {\doibase
  10.1103/PhysRevB.105.094435} {\bibfield  {journal} {\bibinfo  {journal}
  {Phys. Rev. B}\ }\textbf {\bibinfo {volume} {105}},\ \bibinfo {pages}
  {094435} (\bibinfo {year} {2022})}\BibitemShut {NoStop}%
\end{thebibliography}
%
%
%merlin.mbs apsrev4-1.bst 2010-07-25 4.21a (PWD, AO, DPC) hacked
%Control: key (0)
%Control: author (8) initials jnrlst
%Control: editor formatted (1) identically to author
%Control: production of article title (-1) disabled
%Control: page (0) single
%Control: year (1) truncated
%Control: production of eprint (0) enabled
%

\end{document}